\documentclass[longauth]{aa} % for the long lists of affiliations 
\usepackage{graphicx}
\usepackage{color}
\usepackage{epsfig}

\usepackage{amssymb, amsmath}
\usepackage{graphicx,natbib}
\usepackage{pdflscape}

\usepackage{longtable}
\usepackage{lscape}
\usepackage[figuresright]{rotating} 

\usepackage{url}
\usepackage{textcomp}

\usepackage{multirow}

\def\ie{\textit{i.e.}}
\def\q1{\object{q1~Eri}}
\def\bp{\object{$\beta$~Pictoris}}
\def\hd{\object{HD~181327}}

\def\mic{\object{AU~Mic}}
\def\d{\mathrm d}
\def\gra{\textsc{GRaTer}}
\def\um{$\mu$m}
\def\gHG{{$\vert g\dma{\mathrm{HG}}\vert$}}

\def\aap{Astronomy \& Astrophysics}

\def\her{Herschel}

\newcommand{\uma}[1]{^{\mathrm{#1}}}
\newcommand{\dma}[1]{_{\mathrm{#1}}}
\newcommand{\bpr}{\beta\dma{pr}}

\newcommand{\pdx}[2]{#1\times10^{#2}}
\DeclareMathAlphabet{\mathpzc}{OT1}{pzc}{m}{it}

\DeclareTextSymbol{\degre}{OT1}{23}
\begin{document}
\authorrunning{J. Lebreton et al.}
\title{An icy Kuiper-Belt around the young solar-type star \hd }
%\subtitle{Detailed dust and gas modeling}
\titlerunning{An icy Kuiper-Belt around \hd }
\author{
     J. Lebreton\inst{1},
     J.-C. Augereau\inst{1},
     W.-F. Thi\inst{1}, 
     A. Roberge\inst{2}, 
     J. Donaldson\inst{3}, 
     G. Schneider\inst{4},
     S. T. Maddison\inst{5},
     F. M\'enard\inst{1}, 	
     P. Riviere-Marichalar\inst{6},
     G.S. Mathews\inst{7}, 
     I. Kamp\inst{8},	
     C. Pinte\inst{1},
     W.~R.~F. Dent\inst{9},	
     D. Barrado\inst{6,10},  
     G. Duch\^ene\inst{1, 11}, 
     J.-F. Gonzalez\inst{12},
     C. A. Grady\inst{13}, 
     G. Meeus\inst{14},	
     E. Pantin\inst{15},
     J. P. Williams\inst{7},
     P. Woitke\inst{16,17,18}	 %, 
}
%__________________________________________________________________
\institute{
% Thi, Menard, Augereau, Duchene, Lebreton, Pinte, Thi, Martin-Zaidi
UJF-Grenoble 1 / CNRS-INSU, Institut de Plan\'etologie et d'Astrophysique de Grenoble (IPAG) UMR 5274, Grenoble, F-38041, France
\and % Roberge, Donaldson
Exoplanets and Stellar Astrophysics Lab, NASA Goddard Space Flight Center, Code 667, Greenbelt, MD, 20771, USA 
\and % Donaldson
Department of Astronomy, University of Maryland, College Park, MD 20742, USA
\and %Schneider
Steward Observatory, The University of Arizona, 933 North Cherry Avenue, Tucson, AZ 85721, USA
\and % Sarah T. Maddison
Centre for Astrophysics \& Supercomputing, Swinburne University, PO, Box 218, Hawthorn, VIC 3122, Australia
\and % Riviere-Marichalar Barrado
LAEX, Depto. Astrof{\'i}sica, Centro de Astrobiolog{\'i}a (INTA-CSIC),
P.O. Box 78, E-28691 Villanueva de la Ca\~nada, Spain
\and % Williams Matthews
Institute for Astronomy, University of Hawaii at Manoa, Honolulu, HI 96822, USA
\and % Kamp, Podio, Aresu
Kapteyn Astronomical Institute, P.O. Box 800, 9700 AV Groningen, The Netherlands 
\and % Dent de Gregorio
ESO-ALMA, Avda Apoquindo 3846, Piso 19, Edificio Alsacia, Las Condes, Santiago, Chile
%\and % Meeus Eiora
\and % Barrado
Calar Alto Observatory, Centro Astron\'omico Hispano-Alem\'an
C/Jes\'us Durb\'an Rem\'on, 2-2, 04004 Almer\'{\i}a, Spain
\and % Duchene
Astronomy Department, University of California, Berkeley CA 94720-3411 USA
\and % Jean-Francois Gonzalez
Universit\'e de Lyon, Lyon, F-69003, France ; Universit\'e Lyon~1, Observatoire de Lyon, 9 avenue Charles Andr\'e, Saint-Genis Laval, F-69230, France ; CNRS, UMR 5574, Centre de Recherche Astrophysique
de Lyon ; \'Ecole Normale Sup\'erieure de Lyon, Lyon, F-69007, France
\and % Grady
Eureka Scientific and Exoplanets and Stellar Astrophysics Lab, NASA Goddard Space Flight Center, Code 667, Greenbelt, MD, 20771, USA
\and % Meeus
Dep. de F\'isica Te\'orica, Fac. de Ciencias, UAM Campus Cantoblanco, 28049 Madrid, Spain
\and % Pantin
CEA/IRFU/SAp, AIM UMR 7158, 91191 Gif-sur-Yvette, France
 %Woitke, Tilling, Philips, Rice
\and SUPA, Institute for Astronomy, University of Edinburgh, Royal Observatory Edinburgh, UK 
\and % Woitke, Poelman
School of Physics \& Astronomy, University of St.~Andrews, North Haugh, St.~Andrews KY16 9SS, UK
\and % Woitke, Wright
UK Astronomy Technology Centre, Royal Observatory, Edinburgh, Blackford Hill, Edinburgh EH9 3HJ, UK
}
\offprints{jeremy.lebreton@obs.ujf-grenoble.fr}
\date{Received 2011 July 15; accepted 2011 December 7}
%
%Context
\abstract{HD 181327 is a young Main Sequence F5/F6 V star belonging to
  the \bp\ moving group (age $\sim$ 12 Myr). It harbors an optically thin belt of circumstellar material at radius $\sim$90 AU, presumed to result from collisions in a population of unseen
  planetesimals.}
%Aim
{We aim to study the dust properties in the belt in great details, and to constrain the gas-to-dust ratio.}
%{We aim to constrain the disk surface density profile to break the degeneracy between the disk geometry and
 % the dust properties, allowing a detailed study of the grains. 
%	 Knowledge of the dust content is a prerequisite for setting constraints on the diffuse gas counterpart of the disk.
%A first attempt to detect atomic gas lines with Herschel is performed to constrain the gas-to dust ratio.}
%Method
{We obtained far-IR photometric observations of \hd\ with the PACS instrument onboard the Herschel Space Observatory\thanks{Herschel is an ESA space observatory with
    science instruments provided by European-led Principal
    Investigator consortia and with important participation from
    NASA.}, complemented by new 3.2 mm observations carried with the ATCA\thanks{The Australian
  Telescope Compact Array is operated by the Australian Telescope
  National Facility (ATNF) managed by CSIRO.} array.
 The geometry of the belt is constrained with newly reduced HST/NICMOS scattered light images that allow the degeneracy between the disk geometry and the dust properties to be broken.
    We then use the radiative transfer code GRaTer to compute a large grid of models, and we
  identify the grain models that best reproduce the Spectral Energy Distribution (SED) through a Bayesian analysis.
	We attempt to detect the oxygen and ionized carbon fine-structure lines with Herschel/PACS spectroscopy, 
providing observables to our photochemical code ProDiMo.}
%Results
	%Obs
{The HST observations confirm that the dust is confined in a narrow belt. 
The continuum is detected with Herschel/PACS completing nicely the SED in the far-infrared. 
The disk is marginally resolved with both PACS and ATCA. 
	A medium integration of the gas spectral lines only provides upper limits on the [OI] and [CII] line fluxes.
	%Model
	We show that the \hd\ dust disk consists of micron-sized grains of porous amorphous silicates and
  carbonaceous material surrounded by an important layer of ice, for a total dust mass of $\sim0.05M\dma{\oplus}$ (in grains up to 1 mm).
	We discuss evidences that the grains consists of fluffy aggregates. %, and that the larger grains may be more tightly bound to the parent belt. 
  The upper limits on the gas atomic lines do not provide unambiguous constraints:
  only if the PAH abundance is high, the gas mass must be lower than $\sim17M\dma{\oplus}$. }
%
% Conclusions
 {Despite the weak constraints on the gas disk, the age of \hd\ and the properties of the dust disk suggest that it has passed the stage of gaseous planets formation. The dust reveals a population of icy planetesimals, similar to the primitive Edgeworth-Kuiper Belt, that may be a source for the future delivery of water and volatiles onto forming terrestrial planets.}
\maketitle
%
%________________________________________________________________
\section{Introduction}
The evolution of planets in the Solar System is intimately connected to the existence of a reservoir of planetesimals in its outer regions.
The depletion of the Kuiper Belt and the main asteroid belt that occurred during the Late Heavy Bombardment not only delivered large amounts of water, volatiles and carbonaceous material onto the inner planets, but it also reduced the later rate of catastrophic impacts onto the Earth, opening the path to the emergence of life \citep{2000M&PS...35.1309M,2010MNRAS.404.1944G,2011Natur.478..218H}.
Although detecting Kuiper Belt-like objects around nearby planetary systems will remain an unachievable goal for the foreseeable future, their collisional erosion produces circumstellar disks of dust responsible for a characteristic excess emission detectable at infrared and sub-millimeter wavelengths. 
Recent studies using the Herschel Space Observatory reveal that as much as $\sim$30\% of nearby Main-Sequence F, G, K stars are surrounded by cold debris rings analogous to the Kuiper Belt (latest Herschel/DUNES Open Time Key Program results, Eiroa et al. in prep).
However, the cold temperature and the large grain sizes make it impossible to detect solid-state features in their spectrum, thus preventing unambiguous identification of the dust composition. We are thus left with model-dependent methods to constrain the dust composition from color and polarimetric measurements, and from the Spectral Energy Distribution (hereafter SED). 

Another uncertainty in planet formation theories lies in the lack of detailed observational constraints for 
the mutual dust and gas dissipation timescales in circumstellar disks, or more precisely, on
how the gas-to-dust mass ratio evolves with time and location in disks
at the very early stages of planet formation.  
The study of stellar
clusters of different ages shows that, statistically, a population of young stars loses
its massive dust disks in only a few million years on
route to the Main Sequence \citep{carpenter05}, although, for individual
objects, the transition from an optically thick to an optically thin
disk is expected to occur more quickly \citep[a million years or
less,][]{Cieza07,Currie2011}. However not much is known about the
characteristic evolutionary timescales for the gas in these disks.
Observing young stars at different evolutionary stages can help elucidate the extent to which gas and dust simultaneously
dissipate in disks. This requires observations of both the dust and
gas components and, at the same time, modeling of the continuum and line emission.

As part of the GASPS (Gas in Protoplanetary Systems) Herschel Open
Time Key Programme, we observed \hd, a young ($\sim$12\,Myr) F5.5V star
located at 51.8\,pc \citep[][Tab.\,\ref{tab:stardisk}]{Holmberg09} belonging to the \bp\
moving group \citep{Zuckerman04,Mamajek04}. 
\hd\ was identified as a debris disk hosting main sequence star through its SED
with a fractional infrared luminosity $L\dma{IR}/L_{\star} =
0.2\%$ \citep{Mannings98}.
\cite{Schneider06}, from NICMOS coronagraphic
observations in scattered light, discovered a ring-like disk of
circumstellar dust located at radius $\sim$ 89 AU from its star. The ring
is inclined by 31.7$\pm$1.6$\degr$ from face-on and shows an apparent
brightness assymetry with respect to the minor axis that is well
explained by a strong directionally preferential scattering.  
No photocenter nor pericenter offset is seen in the ring relative to the position of
the central star and the disk appears axisymmetric.
%allowing to azimuthally median the surface brightness of the
%deprojected ring. 
Recent reprocessing of these data (Schneider et al., in prep.)
confirms the annulus shape of the disk, and suggests a narrow
radial distribution of the dust in the system (FWHM = 24.5 AU).
%The total flux density of the light scattered by the disk
%corresponds to 0.111$\pm$0.015\% of the starlight at 1.1 \um\ (from $\sim$60 to 200 AU).
%A maximum of $\sim0.76$ mJy.arcsec$^{-2}$ is reached 
%at a radial distance of 88.6 AU with an intrinsic FWHM of 24.5 AU. 
%
A low surface brightness diffuse halo is seen in the NICMOS image and in
complementary HST/ACS images at a distance of $\sim 4'' \le r \le 9''$ from the central star,
which may correspond to a population of very weakly bound grains originating
in the main ring.
%\citet{Schneider06} performed deeper 0.6\,\um\ ACS
%PSF-subtracted coronagraphic observations revealing a faint outer
%nebulosity from 4'' $\le r \le 9$''. 
Mid-infrared images at $18.3$\,\um\ by \citet{Chen08} suggest azimuthal
asymmetries in the density profile of the ring with respect to
the apparent minor axis.
%, which may have lead to an overestimation of the anisotropic
%scattering properties of the grains.  
These $18.3$\,\um\ observations also show that little to no emission
($\lesssim\,15\%$) comes from inside the ring resolved in scattered
light. Therefore, the density profile of the \hd\ dust disk should be 
well constrained by the scattered light and thermal
emission images, allowing a detailed study of the dust
properties. Interestingly, from IRS and MIPS-SED data, \citet{Chen08} also inferred the presence of crystalline ice in the debris disk.

In this paper we perform SED and
line fitting of new Herschel/PACS observations of \hd\ (presented in
Sec.~\ref{Observations}), supplemented by data from the literature. 
We also present new ATCA imaging of the disk (Sec.~\ref{Observations}).
In Sec.~\ref{setup}, we derive surface density profiles from the newly
reduced HST/NICMOS data, which well constrain the ring geometry. 
We describe our dust model and present the results of the SED
fitting in Sec.~\ref{setup} and \ref{sec:dust}.
We assess the gas content in this young debris disk in
Sec.~\ref{sec:gas}. Finally we discuss the results in Sec.\,\ref{sec:discu}, \ref{sec:timescales} and \ref{sec:discuconclu} and present our conclusions in Sec.~\ref{sec:conclu}.
%
%=============== STAR and DISK PROPERTIES =========================
%
\begin{table}[h!btp]
%\begin{center}
  \caption{Star and disk properties}
\label{tab:stardisk}
\begin{tabular}{lcc}
  \hline
  Parameter & Value &  Reference \\
  \hline
  Spectral Type  & F5/F6V &  \citet{nordstrom04} \\
  Magnitude V & 7.0 &  \citet{2000PASP..112..961B} \\
  Gravity (log($g$)) & 4.510 & \citet{Chen06} \\
  Distance ($d_{\star}$) & $51.8$\,pc &  \citet{Holmberg09} \\
  \multirow{2}*{Age} & \multirow{2}*{12 Myr} & \citet{Zuckerman04}\\
  & &	\citet{Mamajek04} \\
  Luminosity ($L\dma{\star}$) & $3.33\,L\dma{\odot}\uma{(a)}$  & this study\\
  Mass ($M\dma{\star}$) &  $1.36\,M\dma{\odot}$  & this study \\
  Disk PA$\uma{(b)}$ & 107$\pm$2$\degr$ & \citet{Schneider06} \\
  Disk inclination & 31.7$\pm$1.6$\degr$ & \citet{Schneider06} \\
  $L\dma{IR}/L\dma{\star}$ & $\sim\pdx{2}{-3}$ & this study\\
    \hline	
\end{tabular}
%\end{center}
%
%
$\uma{(a)}$From integrating the synthetic stellar spectrum, 
$\uma{(b)}$Position Angle, East of North.
\end{table}
%================================================================     
%
%
%________________________________________________________________
\section{Observations and Data Reduction}
\label{Observations}
\subsection{Herschel/PACS continuum}
We obtained new far-IR photometry and spectroscopy of \hd\ using the
PACS instrument onboard the \her\ Space Observatory \citep{Pilbratt10,
  Poglitsch10}. The observations were carried out in two modes: 1)
159~sec of point-source chop-nod mode imaging at 70 and 160~$\mu$m
(obsid: 1342183658) and 2) two 276~sec scan map imaging observations
at 100 and 160~$\mu$m (obsids: 1342209057-8). The scan maps were
executed with the medium scan speed ($20\arcsec /$s), at scan angles
of 70\textdegree\ and 110\textdegree.  The two cross scan maps at
different scan angles were combined to increase the signal-to-noise by
cutting down on streaking in the image background along the scan
direction.

We reduced the photometry with the \her\ Interactive Processing
Environment \citep[HIPE,][]{Ott10} version 4.2. The uncertainties in
the absolute flux calibration for this version are 10\% at 70\, and
100~$\mu$m and 20\% at 160~$\mu$m (see the PACS Scan Map Astronomical Observation Templates release
note\footnote{available at
  \url{http://herschel.esac.esa.int/AOTsReleaseStatus.shtml}}).  For
the 70" and 160~$\mu$m chop-node images, we performed aperture photometry using
apertures of $7\arcsec$ and $11\arcsec$, respectively. For the 100 and 160~$\mu$m scan map
observations, we used apertures of radii $12\arcsec$ and $24\arcsec$,
respectively. An annulus for sky background estimation was placed from
$10\arcsec - 20\arcsec$ beyond the aperture in each case, and an
aperture correction was applied to the measured fluxes. 
Both estimations at 160~$\mu$m are fully compatible within the error bars; the 160~$\mu$m scan-map flux is used in the rest of the study.
Since the source is marginally resolved at 70 and 100~$\mu$m (see
Section~\ref{sub:resolve}), we calculated our own aperture corrections
following the procedure described in \citet{Poglitsch10}.  The
photometry results are presented in
Table~\ref{tab:hd181327_phot_results}. 

\begin{table}[h!t] %\centering
\caption{Photometry results for \hd\ from chop-nod$^{(1)}$ and scan map$^{(2)}$ observations. \label{tab:hd181327_phot_results}}
\begin{tabular}{lcc}
\hline
\multicolumn{1}{c}{Wavelength} & $F_\mathrm{cont}^{\dag}$ & Absolute Flux Cal.\ \\
\multicolumn{1}{c}{($\mu$m)}      & (Jy) & Uncertainty \\
\hline \hline
70$^{(1)}$  & $1.827 \pm 0.0069$ & 10\% \\
100$^{(2)}$ & $1.337 \pm 0.0082$ & 10\% \\
160$^{(1)}$ & $0.767 \pm 0.015$  & 20\% \\
160$^{(2)}$ & $0.772 \pm 0.011$  & 20\% \\

\hline
\multicolumn{3}{l}{\parbox{3in}{\small \vspace*{0.5ex} $^{\dag}$The $\pm 1 \sigma$ error bars on the flux measurements include statistical uncertainty only. }}
\end{tabular}
\end{table} 

\begin{figure}[!hb] \centering
\hspace*{-0.2cm}
\epsfig{file=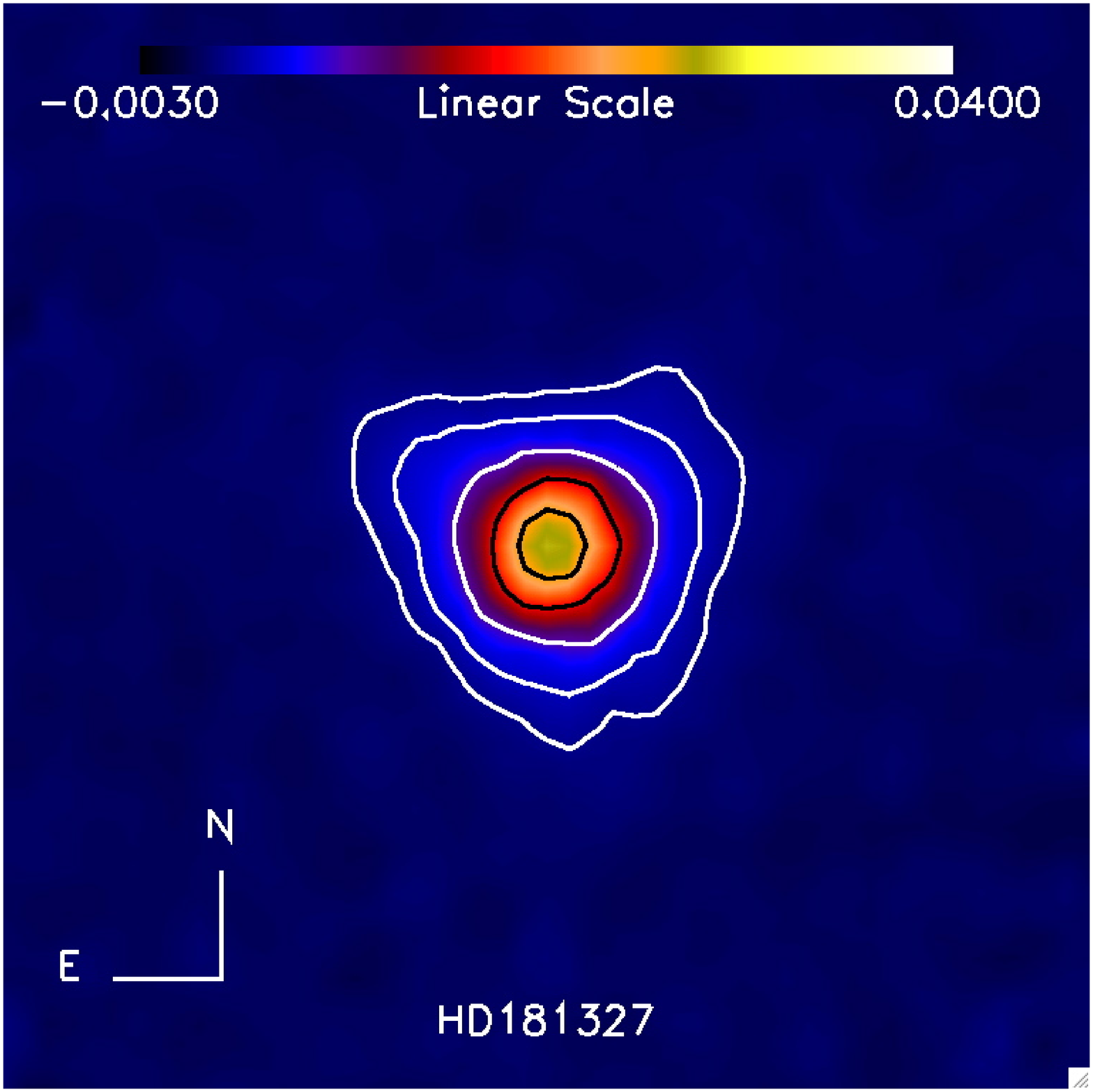, scale=0.16} %\hspace*{0.5in}
\epsfig{file=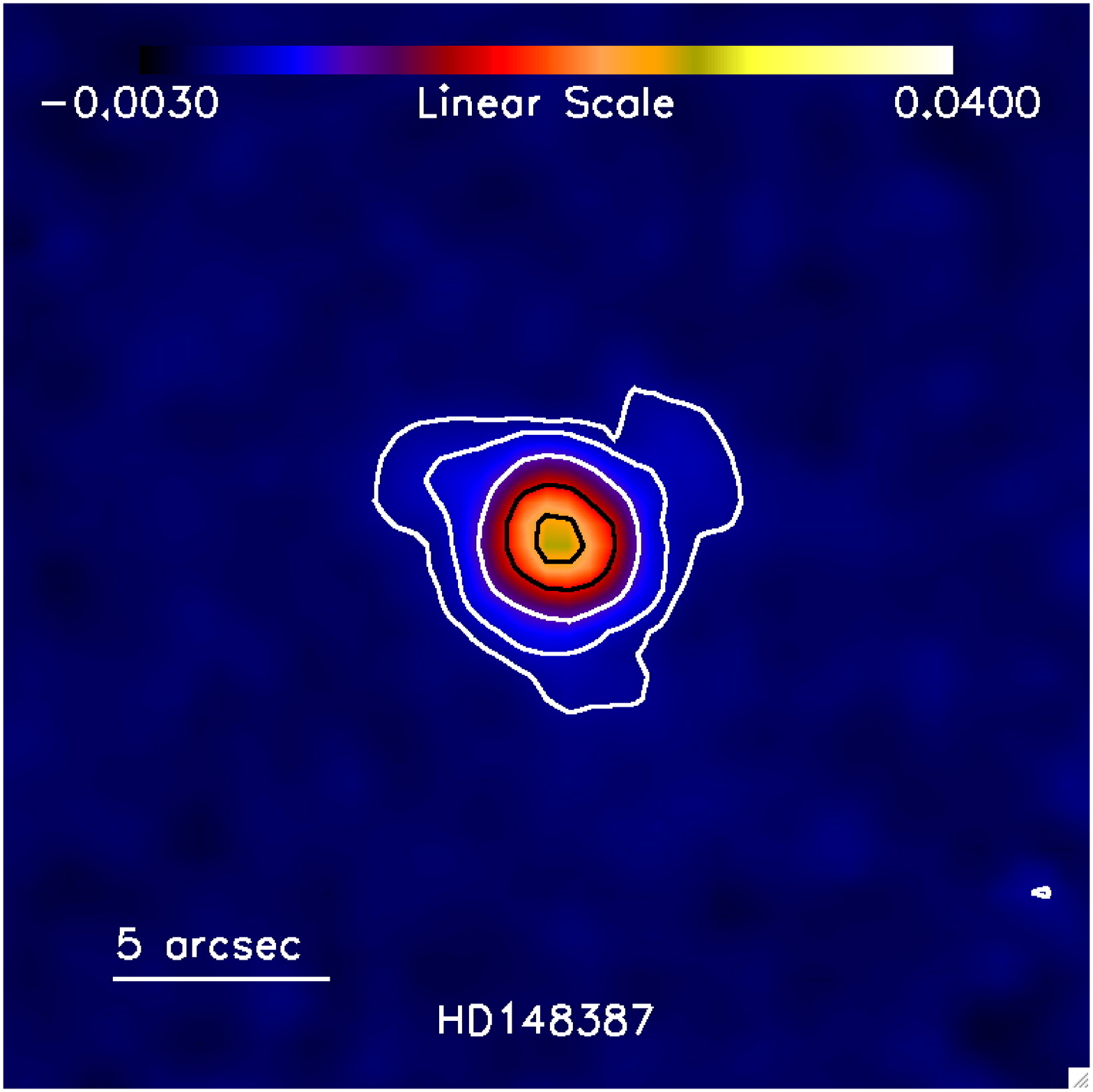, scale=0.16}
\caption{PACS 70~$\mu$m imaging of {\bf (left)} \hd\ and {\bf (right)}
  a PSF reference star, \object{HD\,148387}. Both images were acquired
  in chop-nod mode.  The field of view of each image is $25\arcsec
  \times 25\arcsec$.  Brightness contours at 5, 10, 25, 50 and 75
  times the background noise in the \hd\ image are overlaid on both
  images.
\label{fig:contours}}
\end{figure}

\begin{figure}[h!t]\centering
\epsfig{file=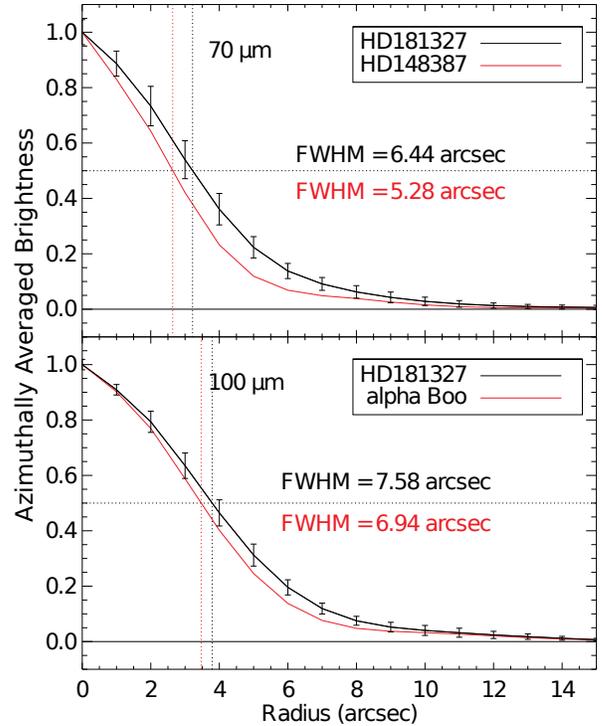, scale=0.35}
\caption{Radial profiles of \hd\ at 70~$\mu$m ({\bf top}), compared
  to \object{HD\,148387} as a PSF reference, and at 100~$\mu$m ({\bf
    bottom}), compared to \object{$\alpha$~Boo} as a PSF
  reference. Both plots show the azimuthally averaged, normalized
  radial brightness profile vs.\ radius from the star. Each error bar
  is the standard deviation of the brightness values at that
  radius. The FWHM values for the profiles are given on the plot.}
\label{fig:rad_prof_70}
\end{figure}

%\subsection{Herschel/PACS imaging}
 \label{sub:resolve}
\hd\ is more extended than the \her\ point-spread function (PSF) in
both the 70~$\mu$m and 100~$\mu$m images.  The 70~$\mu$m image of
\hd\ appears in Figure~\ref{fig:contours}, along with the 70~$\mu$m
image of a PSF reference star (\object{HD\,148387}) observed using the same chop-nod mode. 
The upper panel of Figure~\ref{fig:rad_prof_70}
shows the azimuthally-averaged radial brightness profiles for \hd\ and
\object{HD\,148387}. In the case of the 100~$\mu$m scan map observation of \hd,
we used a different PSF reference observation, i.e.\ a scan map of
\object{$\alpha$~Boo}.  The 100~$\mu$m \object{$\alpha$~Boo} radial profile appears in
Figure~\ref{fig:rad_prof_70}, bottom panel. At both wavelengths, the azimuthally averaged
radial profile of \hd\ is clearly extended beyond that of the PSF
reference.  The FWHM of the \hd\ 70~$\mu$m profile is $6\farcs44$,
compared to a FWHM of $5\farcs28$ for \object{HD\,148387}.  At
100~$\mu$m, the FWHM for \hd\ is $7\farcs58$, compared to $6\farcs94$
for \object{$\alpha$~Boo}.

\subsection{ATCA 3.2\,mm observations}
\label{sec:atca}
The Australia Telescope Compact Array (ATCA) was used to
observe \hd\ at 3.2~mm in May and September 2010.  The observations
were carried out in continuum double sideband mode, with
$2046\times1$~MHz channels per sideband with an effective total
bandwidth of just over 2~GHz. The ATCA has five antenna with 3-mm
receivers, and our observations were carried out with the compact H75
array configuration with baselines of 31--98~m.  The quasars
PKS~1815-553 and PKS~1253-055 were used for the gain and bandpass
calibrations respectively, and Uranus was used for the absolute flux
calibration.  The ATCA calibration uncertainty is estimated to be
about 20\% in the 3-mm band.
\begin{figure}[h!t]
  \begin{center}
    \includegraphics[angle=-0,width=0.95\columnwidth,origin=bl]{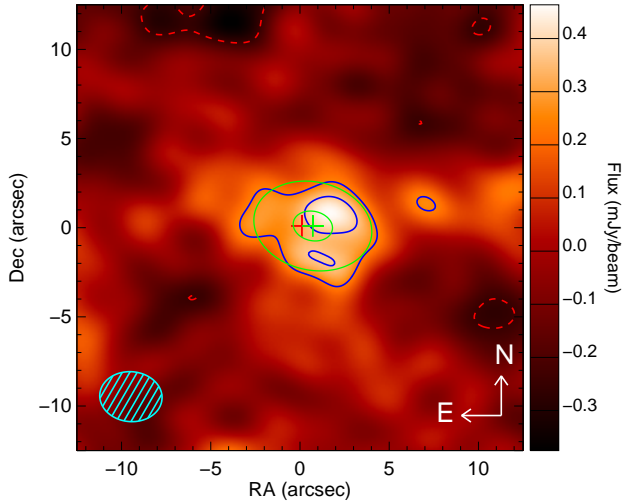}
     \caption{ATCA 93.969 GHz (3.192\,mm) map of \hd\ with contours at -2 (dashed red line), 2 and 3$\sigma$ (solid blue lines), where $\sigma$ is the image RMS given by 1.1$\times10^{-4}$ Jy/beam. The dashed blue area represents the beam. The Gaussian fit is overplot with contours at 2 and 3$\sigma$ (solid green lines). The red and green crosses denote respectively the star position and the Gaussian fit center.}
    \label{fig:atca}
  	\end{center}
\end{figure}
The data were calibrated and reduced with the MIRIAD package \citep{1995ASPC...77..433S}.  The frequency pair for the
observations were centered at 93 and 95~GHz, and during the reduction
process the data was averaged over 64 channels to improve the
signal-to-noise, shifting the frequency pairs to 92.968 and
94.969~GHz. The final CLEANed map combined the two sidebands,
resulting in an image at 93.969~GHz (3192.5~\um) presented in Fig.~\ref{fig:atca}.
Our total on-source integration time was 770~minutes. We determined an
integrated flux of $7.2\times10^{-4}$~Jy, with an RMS of
$\sigma=1.1\times10^{-4}$~Jy/beam. Including the 20\% calibration uncertainty, this yields an uncertainty of 2.5$\times10^{-4}$Jy.
The disk is marginally resolved.
Assuming a Gaussian profile, 
we find a PA of $\sim$\,102\textdegree, consistent with that seen on the NICMOS image ($107\pm2$\textdegree, Tab.\,\ref{tab:stardisk}), and along which the
%FWHM along the semi-major axis (PA\,$\simeq$\,102\textdegree) 
FWHM is $\sim7.8\arcsec$ (beam: $3.5\arcsec\times2.8\arcsec$, {PA~=~85.4\textdegree}).
The center of the Gaussian fit is also shifted by 0.6\arcsec\ toward the West with respect to the star position.
Structures consistent with a belt at $\sim$100 AU 
are detected along the semi-minor axis with a peak flux of $4.01\times10^{-4}$~Jy/beam on the NW side. 
We note a possible brightness asymmetry with respect to the SE side.

\subsection{Herschel/PACS Spectroscopy}

We obtained far-infrared spectroscopy using the PACS Integral Field Unit which provides a 5 $\times$ 5 array of $9\farcs4 \times 9\farcs4$
spaxels. The data were taken in chop-nod mode using the line scan and
range scan modes.  The spectra cover eight small wavelength ranges ($\sim
1.8 \ \mu$m wide) targeting particular emission lines, including [\ion{O}{I}] at
63\,$\mu$m and [CII] at 158\,$\mu$m.  The spectral resolution of the
data varies between $R \approx 1100$ and 3400.  Additional details on
the observations appear in \citet{mat10}.

\begin{figure}[h!btp]
  \begin{center}
    \includegraphics[angle=0,width=0.95\columnwidth,origin=bl]{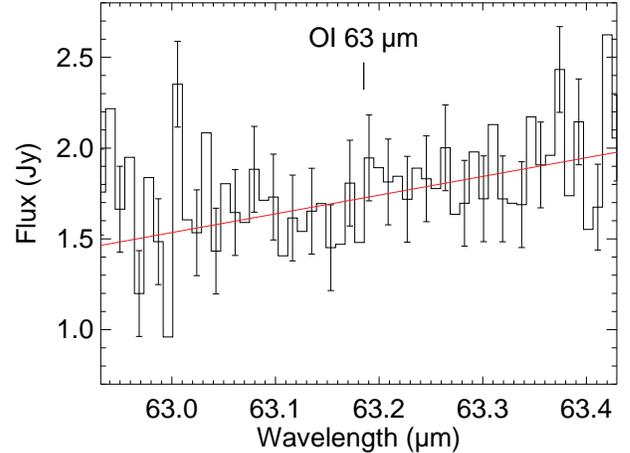}
    \caption{PACS spectrum of HD181327, centered on the wavelength of the
      [\ion{O}{I}] 63~$\mu$m emission line.  The linear
      fit to the continuum flux is overplotted with the red line.  The
      [\ion{O}{I}] 63~$\mu$m emission line was not detected.}
    \label{fig:OI}
  \end{center}
\end{figure}

The data were calibrated with the Herschel Interactive Processing Environment (HIPE) version 4.2.  The wavelength-dependent
flux and aperture corrections provided by the PACS development team
were applied, giving an uncertainty in the absolute flux calibration
of 30\% (see the PACS Spectroscopy Performance and Calibration
Document\footnotemark[\value{footnote}]).
%\footnote{available at
%  \url{http://herschel.esac.esa.int/AOTsReleaseStatus.shtml}}).
Significant emission was detected only in the central spaxel.  No
emission lines are seen, only continuum flux.  The central spaxel
spectrum centered on the [\ion{O}{I}] 63\,\um\ line appears in
Figure~\ref{fig:OI}.
%\begin{figure}[!ht] \centering
%\epsfig{file=OI_63.eps, scale=0.5}
%\caption{PACS spectrum of HD181327, centered on the possible \ion{O}{1} 63~$\mu$m emission line. 
%The 1st degree polynomial fit to the continuum flux is overplotted with the red line.
%The \ion{O}{1} 63~$\mu$m emission line was not detected.  \label{fig:OI}}
%\end{figure}

The continua were binned at half the instrumental FWHM and fitted with 1st degree polynomials, using the fluxes
and errors produced by our HIPE reduction script. 
The HIPE uncertainties
appeared to be slightly too small, however.  New flux random errors were
calculated from the RMS of the data minus the continuum fits, over
wavelength ranges spanning 6 to 14 times the instrumental FWHM,
centered on the expected line rest wavelengths.  The continuum fluxes
at the rest wavelengths of the expected emission lines appear in
Table~\ref{tab:hd181327_spec_results}.  To determine the upper limit
on the integrated line flux ($F\dma{int}$) for each available emission line, we
integrated the data over a small wavelength range, either $\pm 1$ or
$\pm 1.5$ resolution elements, centered on the rest wavelength of the
expected emission line.  The flux errors were propagated to determine
the $1 \sigma$ upper limit on each integrated line flux.  The $3
\sigma$ integrated line flux upper limits appear in
Table~\ref{tab:hd181327_spec_results} as well.

\begin{table}[h!b] %\centering
\caption{Spectroscopy Results for \hd \label{tab:hd181327_spec_results}}\label{tab:spectro}
\begin{tabular}{lrcc}
\hline
Line & \multicolumn{1}{c}{Wavelength} & $F_\mathrm{cont}^{\dag}$ & $F_\mathrm{int}^{\ddag}$ \\
  & \multicolumn{1}{c}{($\mu$m)}      & (Jy) & ($\times 10^{-18} \ \mathrm{W}/\mathrm{m}^2$) \\
\hline \hline
[\ion{O}{I}] & 63.1852  & $1.731\pm0.236$ & $<9.82$ \\
CO\,36-35  & 72.8429  & $1.491\pm0.185$ & $<18.41$ \\
o-H$_2$O    & 78.7414  & $1.431\pm0.187$ & $<10.51$ \\
CO\,29-28  & 90.1630  & $1.264\pm0.222$ & $<8.90$ \\
{[\ion{O}{I}]} & 145.5350 & $0.804\pm0.107$ & $<8.48$ \\
{[\ion{C}{II}]} & 157.6790 & $0.767\pm0.175$ & $<7.02$ \\
o-H$_2$O    & 179.5265 & $0.388\pm0.201$ & $<6.87$ \\
DCO$^+$\,22-21    & 189.5700 & $0.581\pm0.378$ & $<16.71$ \\
\hline
\end{tabular}
%\multicolumn{4}{l}{\parbox{3in}{\small \vspace*{0.5ex} $^{\dag}$ The $\pm 1
%    \sigma$ error bars on the flux measurements include statistical
%    uncertainty only. The absolute calibration uncertainty amounts
%    to 30\%. $^{\ddag}$ $3\sigma$ upper limit }}
$^{\dag}$ The $\pm 1 \sigma$ error bars on the flux measurements
include statistical uncertainty only. The absolute calibration
uncertainty amounts to 30\%. $^{\ddag}$ $3\sigma$ upper limit
\end{table} 

%
%\begin{figure}[!ht] \centering
%\epsfig{file=OI_63.eps, scale=0.5}
%\caption{PACS spectrum of HD181327, centered on the possible \ion{O}{1} 63~$\mu$m emission line. 
%The 1st degree polynomial fit to the continuum flux is overplotted with the red line.
%The \ion{O}{1} 63~$\mu$m emission line was not detected.  \label{fig:OI}}
%\end{figure}
%
%________________________________________________________________
\section{Model Setup}
\label{setup}
The dust and gas models developed in Secs.\,\ref{sec:dust} and
\ref{sec:gas} rely on some assumptions on the spatial distribution of
the dust and the gas, as well as on the dust properties. These are
detailed below in Secs.\,\ref{sec:RadialProfile} and
\ref{sec:dustproperties}. Additional data collected from the literature
and stellar properties are summarized in Sec.\,\ref{sec:adddata}.

\subsection{New Surface Density Profile for the \hd\ Ring}
\label{sec:RadialProfile}
Our spatially resolved imaging constrains the location of the dust.  We advantageously impose this constraint in order to examine the dust properties in details and 
to get an upper limit
on the gas mass from the \her\ non-detections of gaseous emission lines. 
The disk around \hd\ was spatially resolved in scattered light with HST/NICMOS
1.1\,\um\ imaging by \citet{Schneider06}, which alleviates many
uncertainties on the disk structure. The NICMOS images revealed
a ring-like disk with a peak density position $r_{0}=1.71$''
\citep[{$\sim$~88.6\,AU}, according to the new distance to the star, 51.8 pc,
  by][]{Holmberg09}, with an azimuthally medianed photometric FWHM of $\sim$ 0.538'' ({$\sim 28$~AU}).

\begin{figure}[h!tbp]
  \begin{center}
    \hspace*{-0.4cm}
    \includegraphics[angle=0,width=1.05\columnwidth,origin=bl]{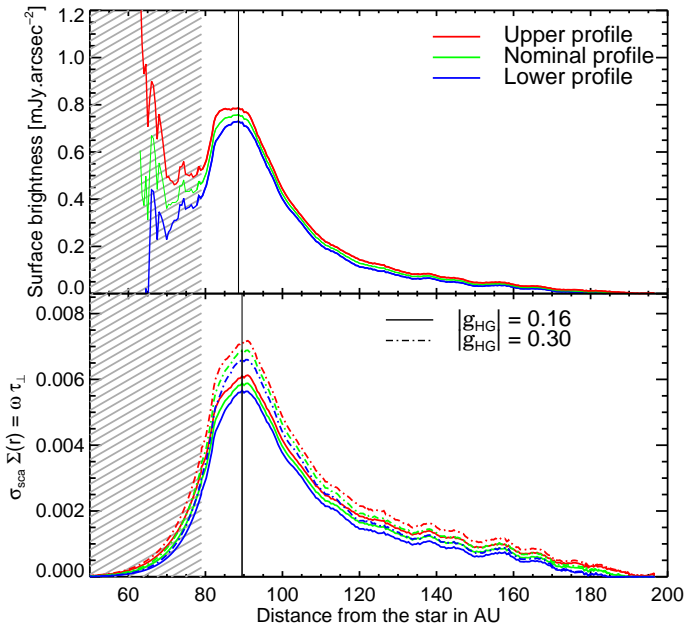}
     \caption{New HST/NICMOS 1.1\,\um\ surface brightness profiles
       ({\bf top}, nominal and $\pm 1\sigma$ profiles), and surface
       density profiles
%($\Sigma(r)$) multiplied by the mean
%       scattering cross section at $1.1\,\mu$m ($\sigma\dma{sca}$)
       resulting from direct inversion of the brightness profiles
       ({\bf bottom}), assuming two possible values for the asymmetry parameter. The color-coding is the same for both panels.
       The striped areas identify regions dominated by PSF-substraction
       residuals where the density has been extrapolated.}
    \label{fig:SDP}
  	\end{center}
\end{figure}

Recently, Schneider et al. (in prep.) improved the data reduction of the
original HST/NICMOS observations of \hd\ 
thanks to PSF subtraction template images better matched to the stellar PSF structure in the HD 181327 images.  This results in an overall improvement of the signal-to-noise, especially inside the ring, and leads to a slight sharpening of
the belt.
% (FWHM of $\sim24$\,AU when extrapolating the inner region below
%$\sim$1.50'', where it is dominated by PSF substraction residuals).
The new profiles, displayed in the top panel of Fig.\,\ref{fig:SDP},
show a peak at a distance of $88.6$\,AU and a surface brightness
fainter by $\sim21\%$ (0.76\,mJy.arcsec$^{-2}$ at the peak position)
compared to the original \citet{Schneider06} study.

We derive the surface density profile $\Sigma(r)$ by
inverting the new scattered light surface brightness profile. We
employ the methodology described in \citet{Augereau06}, assuming a
disk inclination of $i=31.7$\degr\ from face-on. The inversion
procedure yields the product $\sigma\dma{sca}\Sigma(r)$, where
$\sigma\dma{sca}$ is the mean scattering cross section of the grains
at 1.1\,\um. This is equivalent to the product of the vertical optical
depth $\tau\dma{\perp}(r)$, multiplied by the mean albedo $\omega$ at
that wavelength \citep[$\sigma\dma{sca}\Sigma(r) =
  \omega\tau\dma{\perp}(r)$, see Eqs.\,2--3 in][]{Schneider06}.
%
%The dust density profile used in this study is derived from a direct
%inversion of the new scattered-light surface brightness profile, using
%the methodology of \citet{Augereau06}.
%% Precise: axisymmetric, gHG
%We obtain the product of the mean scattering cross section of the
%grains at 1.1 \um\ $\sigma\dma{sca}$, and the disk surface density
%$\Sigma(r)$ (in the cylindrical frame centered on the star and
%attached to the disk), by inverting Eq.\,2 of \citet{Schneider06} and
%using their Eq.\,3.  This quantity, also equivalent to the vertical
%optical depth $\tau\dma{\perp}$, multiplied by the mean albedo
%$\omega$ ($\sigma\dma{sca}\Sigma(r) = \omega\tau\dma{\perp}$),
%requires no \textit{a priori} assumption on the grain properties.  
%
The inversion solution depends on both the anisotropic
scattering phase function, and on the disk vertical profile.
Nevertheless, \citet{Schneider06} showed that the density profile is
largely independent of the shape for the vertical dust
distribution in the range of acceptable vertical scale heights $H_0$.
%, and it shall not depend severely on the exact vertical
%profile, as long as the scale height, $H_{0} = H(r = r_{0})$, is
%sufficiently large.  We use a typical Gaussian and linear profile
%%($\gamma = 2$, $\beta = 1$)
We here assume a Gaussian vertical profile with a scale height $H_{0}
= 6$\,AU at the observed peak radius of the ring $r_{0} = 88.6$\,AU
\citep[$H_0/r_0 \sim 0.068$, consistent with the theoretical lower
  limit of 0.04 from][]{Thebault09}.  The observed phase function of
the disk at 1.1\,\um\ is well-represented by a Henyey-Greenstein
scattering phase function with asymmetry parameter \gHG$ =
0.30\pm0.03$. This value was discussed by \citet{Chen08} in light of
asymmetries identified in T-ReCS thermal emission images in the
$Q_{a}$-band.  \citet{Chen08} suggest that the asymmetries may have
led to an overestimate of \gHG, which may be as small as
$0.16\pm0.04$.
%
%In that approach, $\Sigma(r)$ only depends on the phase function in
%the range of scattering angles probed by the observations (between
%about $\pi/2-i$ and $\pi/2+i$). Assuming the dust is homogeneously
%distributed in azimuth, the disk geometry (after sky plane
%deprojection) is well represented by a Henyey-Greenstein scattering
%phase function with asymmetry parameter \gHG$ = 0.30\pm0.03$ at 1.1
%\um.  If however, the bilateral asymmetry with respect to the
%semi-major axis seen in the NICMOS scattered light image, and possibly
%in T-ReCS thermal light image in the $Q_{a}$-band, corresponds to a
%true asymmetry of the dust density, then \citet{Chen08} propose a
%revised \gHG$ = 0.16\pm0.04$ corresponding to a smaller forward- or
%backward-scattering effect.  
%
We thus perform the inversion for both \gHG $= 0.16$ and \gHG =
$0.30$.
%In order to anticipate the possible consequence of having large
%uncertainties in the inner parts of the disk, we used three distinct
%radial brightness profiles : a nominal brightness profile, defined as
%the measured profile, an upper profile defined by the nominal profile
%plus the 1$\sigma$ error bars, and a lower profile defined by the
%nominal profile minus the 1$\sigma$ error bars (Fig.\,\ref{fig:SDP},
%top panel).  This leads to a total of 6 radial density profiles.

The results are displayed in the bottom panel of Fig.\,\ref{fig:SDP},
where we also show the impact of the surface brightness
uncertainties on the inferred surface density profile. We find an
actual surface density peak at 89.5\,AU from the star.  A higher
\gHG\ leads to a globally higher surface density profile. 
The surface brightness measured
interior to $\sim1.5$'' ($\sim78$ AU) is not reliable because of PSF
subtraction residuals. We fit the profiles from 79 to 83\,AU obtained
for both \gHG\ values by a very steep power-law function of the form $\sim
r^{+10\pm1}$. We extrapolate this profile interior to 78 AU as shown in
Fig.\,\ref{fig:SDP} (striped area). This extrapolation is supported by the mid-IR results of
\citet{Chen08}, who show that the inner part of the disk is largely
cleared toward the center, containing little to no dust. 

Outside of the belt peak position, the improved data reduction results in a profile that falls off as $\sim r^{-4.7\pm0.3}$ from $\sim$92 to 162 AU, a steeper fall-off than the
\citet{Schneider06} original profiles ($\sim r^{-3}$). 
The dust is thus more tightly confined to the parent-belt than previously estimated, and the profile is also steeper than theoretical expectations.
The exact value of the outer slope can only slightly impact the modelling results, as most of the emission arises from the inner part of the belt where the temperature is larger, except possibly for the longer-wavelengths.

Eventually, the differences between the several radial profiles in Fig.\,\ref{fig:SDP} essentially
translate into a scaling factor that impacts the total disk mass
estimate. We estimate that the uncertainty on the asymmetry parameter adds an additional $\sim$4\% to the uncertainty on our final mass values. 
The density profile obtained for \gHG=0.30 is used in
the rest of this study. We stress that, in order to focus on the other disk properties, we make the assumption that this profile is valid for any dust grain constrained by the SED, \ie\ for grain sizes ranging from the micron to the millimetre scale.
This assumption assumes no radial segregation in dust sizes and it remains reasonable as long as the flux at any wavelength arises predominantly from regions close to the density peak position
(see Sec.\,\ref{sec:discusslongwav}).
%
%Estimated from the peak density position, we will have to add an
%uncertainty of $\sim$ 3.7\% to the mass uncertainties we obtain.
%
%=============== PHOTOMETRY TABLE  ============================================
%
%
\begin{table}[!hbtp] % \centering
  \caption{\hd\ photometric observations from the literature. The
    \her\ continuum values used for the dust modeling are in
    Tables \ref{tab:hd181327_phot_results} and
    \ref{tab:hd181327_spec_results}. \label{tab:photometry}}
\begin{tabular}{cccr}
\hline
Wavelength  & Flux & Uncertainty & Instrument \\ 
($\mu$m)    & (Jy) & (Jy)        & \& Reference \\ 
\hline
12	&	0.114	&	0.011	&	IRAS [1]\\
15.0	&	0.094	&	0.004	&	Spitzer/IRS [2]\\
16.3	&	0.097	&	0.005	&	Spitzer/IRS [2]\\
17.4	&	0.107	&	0.005	&	Spitzer/IRS [2]\\
18.3	&	0.114	&	0.003	&	T-ReCS	[3]\\
18.7	&	0.125	&	0.007	&	Spitzer/IRS [2]\\
19.8	&	0.144	&	0.009	&	Spitzer/IRS [2]\\
20.4	&	0.149	&	0.008	&	Spitzer/IRS [2]\\
21.1	&	0.157	&	0.020	&	Spitzer/IRS [2]\\
23.1	&	0.224	&	0.007	&	Spitzer/IRS [2]\\
23.7 &	0.223	&	0.009	&	Spitzer/MIPS [1]\\
25	&	0.306	&	0.024	&	IRAS [1]\\
25.7	&	0.303	&	0.014	&	Spitzer/IRS [2]\\
28.0	&	0.384	&	0.014	&	Spitzer/IRS [2]\\
30.6	&	0.488	&	0.026	&	Spitzer/IRS [2]\\
33.1	&	0.623	&	0.025	&	Spitzer/IRS [2]\\
35.5	&	0.718	&	0.034	&	Spitzer/IRS [2]\\
54.5    &	1.88	&	0.09	&	MIPS-SED [4]\\
58.0	&	1.83	&	0.07	&	MIPS-SED [4]\\
60	&	1.73	&	0.17	&	ISO	 [5]\\
61.4	&	1.98	&	0.08	&	MIPS-SED [4]\\
64.9	&	2.01	&	0.08	&	MIPS-SED [4]\\
68.4	&	1.98	&	0.08	&	MIPS-SED [4]\\
%70	&	1.827	&	0.1828	&	PACS70 [6] \\
71.4	&	1.73	&	0.12	&	Spitzer/MIPS [1]\\
71.6	&	1.99	&	0.09	&	MIPS-SED [4]\\
75.1	&	1.93	&	0.10	&	MIPS-SED  [4]\\
78.4	&	1.81	&	0.10	&	MIPS-SED  [4]\\
81.8	&	1.72	&	0.11	&	MIPS-SED  [4]\\
85.3	&	1.62	&	0.11	&	MIPS-SED  [4]\\
88.8	&	1.47	&	0.11	&	MIPS-SED  [4]\\
90	&	1.41	&	0.14	&	ISO	[5]\\
92.0	&	1.46	&	0.13	&	MIPS-SED  [4]\\
95.5	&	1.37	&	0.17	&	MIPS-SED  [4]\\
100	&	1.7	&	0.2	&	IRAS [1]\\
%100	&	1.337	&	0.1339	&	PACS100 [6]\\
155.9	&	0.77	&	0.09	&	Spitzer/MIPS[1]\\
%160	&	0.772	&	0.1548	&	PACS160 [6]\\
170	&	0.736	&	0.192	&	ISO [5]\\
870	&	0.052	&	0.006	&	LABOCA	[6]\\
3190    &     7.2$\times 10^{-4}$  &   2.5$\times 10^{-4}$  &       ATCA [7] \\
\hline 
\end{tabular}
{\sc Notes and References --} All uncertainties are $1\sigma$ and include absolute calibration error. 
[1] \citet{Schneider06}, 
[2] This study, data reduced with the Spitzer/c2d team pipeline \citep{c2d},
[3] Gemini, \citet{Chen08}, 
[4] Spitzer, \citet{Chen06},
[5] \citet{Moor06},
%[6] This study. Uncertainties include both statistical error and
%absolute 10\% calibration error added in quadrature at 70 and 100\,\um,
%and 20\% at 160\,\um,
[6] APEX, \citet{Nilsson09},
[7] This study, Sec.\,\ref{sec:atca}.
\end{table}
\subsection{Spectral Energy Distribution}
\label{sec:adddata}
%
%\subsubsection{Data from the literature}
The \her\ and ATCA data presented in this paper are complemented by IR and
sub-mm data
%measured by several instruments and 
collected from the literature.  These are listed in Table
\ref{tab:photometry}.  The shortest wavelength flux we use is the
$12$\,\um\ IRAS photometry for which the emission shows no excess from
the stellar photosphere level.  This observation is documented in
\citet{Schneider06} together with IR-excess from the IRAS Faint Source
Catalogue and obtained with the Spitzer/MIPS instrument.  These are
complemented by ground-based mid-IR measurements performed with the
T-ReCS imager on Gemini-S by \citet{Chen08}, and 13 points extracted
from the 5--35\,\um\ Spitzer/IRS spectrum that we reduced using the c2d
Legacy team pipeline \citep{c2d}.

Additional MIPS observations were obtained in SED-mode by
\citet{Chen06} between 54.5 and 95.5 \um, well-constraining the shape
of the disk SED around its maximum emission wavelength.  Far-IR excess
was reported by \citet{Moor06} using the ISO database. 
The shape of the SED in the far-IR domain (100-200 $\mu$m) is well constrained by our new Herschel measurements.

\begin{figure}[btp]
  \begin{center}
    \includegraphics[angle=0,width=\columnwidth,origin=bl]{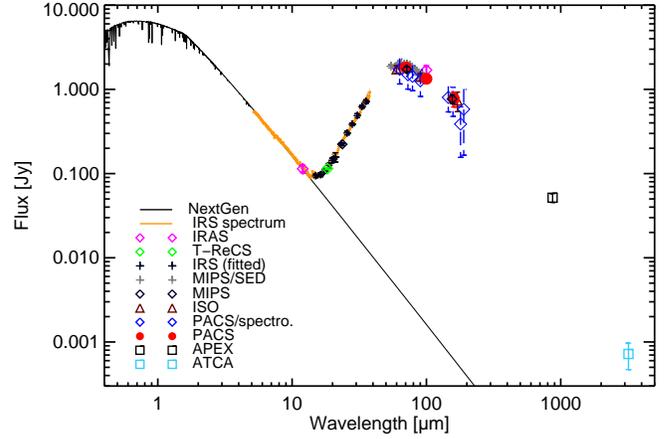}
    \caption{Observed Spectral Energy Distribution of \hd\ (see
      Tables\,\ref{tab:hd181327_phot_results},
      \ref{tab:hd181327_spec_results} and \ref{tab:photometry})}
    \label{fig:ObservedSED}
  \end{center}
\end{figure}

Because the uncertainties on the dust optical constants at millimetre wavelengths are high, we chose 
not to use the ATCA flux to constrain the dust model (see Sec.\,\ref{sec:discusslongwav} for a discussion on the compatibility of the model to these observations).
%However,
%we could not consider the ATCA flux to fit the SED, due notably to the
%lack of optical indexes for the ACAR sample of amorphous carbon beyond
%$\sim$2\,mm. 
Therefore, the sub-mm
regime is only represented by a measurement with the LABOCA bolometer
array at the 12-m telescope APEX \citep[][{$\lambda \sim 870$\,\um}]{Nilsson09},
%, and the millimeter
%regime by our recent ATCA measurement (Sec.\,\ref{sec:atca}). 
%Therefore, the LABOCA measurement 
which thus represents a key piece
of information. 
%The consistency of our modeling
%results with the ATCA measurement is discussed in
%Sec.\,\ref{sec:discusslongwav}. 
These archival data, together with our new observations, constitute the 49 measurements that are displayed in
Fig. \ref{fig:ObservedSED}.
%
%=============== SED =========================
%
\subsection{Star Properties and dik geometry}
We derive excesses from a comparison of the SED with a
synthetic stellar spectrum extracted from the ``NextGen'' model
atmosphere grid \citep{nextgen} and scaled to match the visible
magnitude V = 7.0 mag \citep{2000PASP..112..961B}, assuming ${\log(g) = 4.5}$, ${T_{\mathrm{eff}}=6600}$K,
which leads to a stellar luminosity of 3.33\,$L_{\odot}$
and a mass of 1.36\,$M_{\odot}$.  This stellar model shows good
agreement with the Spitzer/IRS spectrum at wavelengths shorter than
10--15\,\um, longward of which the excess starts dominating over the
stellar photospheric emission.  The star properties and disk geometry adopted in this
study are summarized in Table\,\ref{tab:stardisk}.

\subsection{Dust Properties: Size Distribution and Composition}
\label{sec:dustproperties}
%
%=============== TABLE  =====================================================
%
\begin{table*}[h!tpb]
\begin{center}
  \caption{Parameter space explored with \gra\,, corresponding to a
    total of about 1,150,000 models. The volume fractions $v\dma{Si}$,
    $v\dma{C}$, $v\dma{ice}$ and the porosity $\mathcal{P}$ are
    defined in Sec.\,\ref{sec:dustproperties} (Eq.\,\ref{eq:volumes})}
\label{tab:graterparameters}
\begin{tabular}{cccccc}
\hline
 &Description & Parameter & Explored range &  Values & Distribution\\ 
\hline
\hline
\multirow{4}*{\it Sec.\,\ref{sec:dustproperties}} &minimum grain size & $a\dma{min}$ [$\mu$m] & 0.01\ldots 100 & 77 & log \\
&size power law index &$\kappa$ & -$6.0$\ldots -$2.5$ & 50 & linear \\
&maximum grain size & $a\dma{max}$ [mm] & $7.8$ & fixed & --  \\
&Disk mass up to 1 mm & $M\dma{dust}$ [$M_{\oplus}$] & $> 0$ & fitted & -- \\
 & & \\
\vspace*{-0.6cm} \\
\hline\hline
& {\sc 1 and 2-material models} & &&&\\
\hline
{\it Sec.\,\ref{sec:si+acar}} &{Si+Amorphous C}  & ACAR volume fraction $v\dma{C}$  & $0.00$\ldots $1.00$ & 21 & linear\\
%\hline
{\it  Sec.\,\ref{sec:astrosi+ice}} &{Si+amorphous ice} & ice volume fraction $v\dma{ice}$  & $0.00$ \ldots $0.90$ & 19  & linear\\
%\hline
{\it Sec.\,\ref{sec:astrosi+ice}} & {Si+crystalline ice}  & ice volume fraction $v\dma{ice}$ & $0.00$ \ldots $0.90$ & 19  & linear\\
%\hline
{\it  Sec.\,\ref{sec:porous}} & {Porous Si} & Porosity $\mathcal{P}$ & $0.00\ldots 0.99375$$\uma{(a)}$ & 24  & linear  \\
\\
\vspace*{-0.6cm} \\
\hline\hline
&{\sc 3 and 4-material models} &&&& \\
\hline
\multirow{4}*{\it Sec.\,\ref{sec:ism}}&Si& astrosi volume fraction $v\dma{Si}$ & 1/3$\times$(1-$v\dma{ice}$) & -- & --   \\
&+amorphous C& am. carbon volume fraction $v\dma{C}$ & 2/3$\times$(1-$v\dma{ice}$) & -- & -- \\
&+amorphous ice& ice volume fraction $v\dma{ice}$  & $0.00$\ldots $0.90$$\uma{(b)}$ & 12  & linear  \\
&+vacuum& Porosity  $\mathcal{P}$ & $0.00$\ldots $0.95\uma{(c)}$ & 18  & linear  \\
\hline
\end{tabular}
\end{center}
\vspace*{-0.3cm}
{\sc Notes -- }
$\uma{(a)}$~19 linearly distributed points from 0.00 to 0.95 + additional high porosity points : 0.96875, 0.975, 0.9875, 0.99375.
$\uma{(b)}$~10 linearly distributed points from 0.00 to 0.90 + additional points at 0.65 and 0.75. 
$\uma{(c)}$~10 linearly distributed points from 0.00 to 0.90 + additional points at 0.45, 0.55, 0.625, 0.65, 0.675, 0.75, 0.85, 0.95.
\end{table*}
%
%
%subsubsection{Surface density distribution}
%We assume a single population of grains distributed radially according
%to the surface density profile derived from inverting the scattered
%light surface brightness profiles observed with the HST in the near-IR
%band. We refer to Sec.\,\ref{Observations} for details about how this
%was derived and uncertainties on the profile. For a given model, the
%total mass of the disk, $M\dma{dust}$, or equivalently, the surface
%density at a reference position, is then scaled to best fit the
%observations reported in Table\,\ref{tab:photometry}.

%\subsubsection{Grain size distribution}
For both the dust and gas modeling, we make the assumption
that the grain population has a single differential size distribution valid
anywhere in the ring. This is a central hypothesis that necessarily limits the uniqueness of the solution.
We discuss its impact in Sec.\,\ref{sec:discu}.
The differential size distribution follows a classical power law $\d n(a)
\propto a^{\kappa}\d a$, from a minimum size $a\dma{min}$ to a
maximum size $a\dma{max}$. The minimum grain size $a\dma{min}$ and the
power law exponent $\kappa$ are free parameters that are constrained
by the SED fitting (Sec.\,\ref{sec:dust}). The maximum grain size
$a\dma{max}$ is a fixed parameter, sufficiently large so
that it does not affect the light emission properties at the
wavelengths considered (we chose $a\dma{max} \sim 8\,\mathrm{mm}$).
%In this paper, we consider a grid of 77 logarithmically distributed
%values of $a\dma{min}$ between 0.01\,$\mu$m and 100\,$\mu$m and 50
%linearly distributed values of $\kappa$ between -6.00 and -2.50.
For standard $\kappa$ values (close to -3.5 or smaller), the mass of solid material in a debris disk can be "arbitrarily" large 
depending on the assumed higher cut-off value of the size distribution (that can extend up to planetesimal sizes).
In this article, we define the dust mass $M_{\mathrm{dust}}$ as the total mass of
material contained in grains smaller than 1 mm.
This definition has been followed within the GASPS consortium because it allows direct comparison between different studies independently from the model assumptions.

\begin{figure}[h!btp]\label{fig:opacity}
\begin{center}
 \hspace*{-0.4cm} \includegraphics[angle=0,width=1.05\columnwidth,origin=bl]{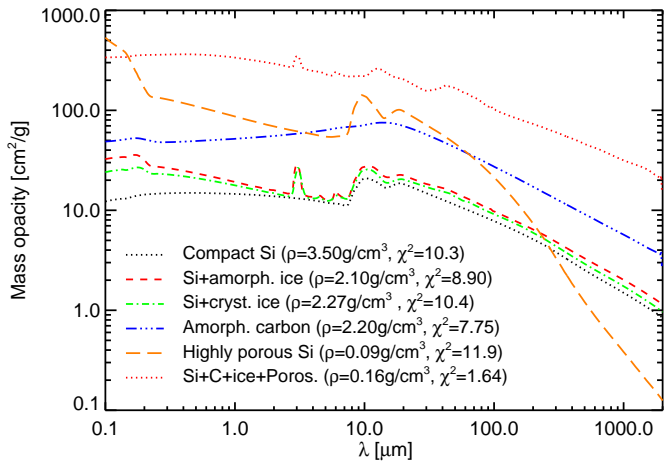}
  \caption{Mass opacity of spherical grains computed with the Mie theory and Bruggeman mixing rule for
 several prototypical compositions detailed in Tab.\,\ref{tab:dustmodels} and their associated best-fit size distributions.}
  \end{center}
\end{figure}
%
%\label{sec:compo}

We adopt a complex grain model that bears some similarities to
those succesfully employed by several authors for fitting SEDs of
young debris disks \citep{LiGreenb,Augereau99a,LiLu03a,LiLu03b}, and
for fitting scattered light disk colors \citep{kohler2008}. We
consider icy porous aggregates, with silicate cores and organic
refractory mantles (or simplified versions of this model; see
Sec.\,\ref{sec:dust}), to mimic fluffy particles made of essentially
unalterred interstellar material \citep[the so-called ``cold-coagulation''
  dust model in][]{LiLu03a,LiLu03b}.
%
%
%It is widely accepted that silicates are the key component of dust
%particles in the interstellar medium.  Cosmic abundances lead to a
%substantial mass fraction of silicates and they are revealed by their
%10\,\um \ and 20\,\um\ emission and/or absorption features detected in
%young circumstellar dust disks \citep[\textit{e.g.}][]{Olofsson2009}.
%However, one can calculate that such grains would be rapidly destroyed
%by photosputerring or collisions. %\footnote{In which environment?
                                %ref} 
%
%\citet{LiGr97} propose that astronomical silicates are likely
%surrounded by a mantle of amorphous carbonaceous material.
%% which
%%preserves the cores from being destroyed faster than they are
%%produced, thus extending the lifetime of the particles.  
Assuming cosmic abundances, the volume of the carbonaceous mantle
$V\dma{C}$ is estimated to be $\sim$ 2.12 times that of the silicate
core $V\dma{Si}$ in the interstellar medium, a value similar to the
one derived from the mass spectra of comet Halley in the Solar System
\citep[$V\dma{C}/V\dma{Si} \sim 2$, ][]{Greenberg98}.

To calculate the optical properties, we use a number of representative
materials with well established optical indexes. These are either
derived from laboratory measurements or from observational and
analytical arguments. The materials are: (1) the astronomical silicates (astroSi) from
\citet{Draine2003}, a theoretical material consisting of amorphous
silicates polluted by carbonaceous materials and other metals
(${\rho=\mathrm{3.5\ g.cm\uma{-3}}}$); (2) the ACAR sample of
amorphous carbon from \citet{zubko} as organic refractories
(${\rho=1.95\ \mathrm{g.cm^{-3}}}$); (3) the H$_{2}$O-dominated ice
from \citet{LiGreenb} (amorphous ice,
${\rho=1.2\ \mathrm{g.cm^{-3}}}$); and (4) the crystalline water ice
from \citet{Warren84} (${\rho=0.92\ \mathrm{g.cm^{-3}}}$).

The optical indexes of an aggregate are obtained using a standard
effective medium theory (EMT), the Bruggeman mixing rule \citep{Bohren83},
and following the methodology of \citet{LiGreenb} and
\citet{Augereau99a}.  The optical properties are computed with Mie
theory valid for hard spheres. The EMT-Mie combination has been shown
to be a good approximation for the absorption and scattering
efficiencies, provided that grain shapes are close to ellipsoids such
that the porosity can effectively be seen as the volume fraction of
vacuum \citep[][and references therein]{Shen2009}. In fact,
\citet{Shen2009} show that only the polarization and the phase
functions are significantly sensitive to grain shapes.
%We apply the Mie approximation and Bruggeman EMT to grains that we
%assume spherical and homogeneous for simplification.
%%
%We do not know however whether ice is homogeneously distributed inside
%the grains and whether grain porosity is well modelled by mixtures of
%matter and vacuum. Keeping in mind that dust particles usually have
%irregular shapes and fluffy structures, it makes little sense to
%consider the ice should accumulate only onto the outer layer of the
%grains.  
%\citet{Shen2009} constructed a model of random ballistic aggregates to
%evaluate optical properties for realistic grain shapes. They conclude
%that the EMT-Mie model is a good approximation for the absorption and
%scattering efficiencies provided that grain shapes are close to
%ellipsoids such that the porosity can effectively be seen as the
%volume fraction of vacuum.  They show that only the polarization and
%the phase functions are significantly sensitive to grain shapes.
%
We define the composition of a grain through the volume fractions of
its different components :
\begin{eqnarray}
	v\dma{Si}  =  \frac{V\dma{Si}}{V\dma{mat}} %	\label{eq:vsi}\\
         , \, v\dma{C} = \frac{V\dma{C}}{V\dma{mat}}  %	\label{eq:vc}\\
         , \, v\dma{ice} = \frac{V\dma{ice}}{V\dma{mat}}, 
        \, \mathcal{P} = \frac{V\dma{vac}}{V\dma{mat}+V\dma{vac}}	\label{eq:volumes}
\end{eqnarray}
where $V\dma{Si}$, $V\dma{C}$, $V\dma{ice}$ and $V\dma{vac}$ are the
total volume of astronomical silicates, carbonaceous material,
%(amorphous or crystalline) 
ice and vacuum in the grain, respectively, and $V\dma{mat} =
{V\dma{Si}+V\dma{C}+V\dma{ice}}$ ~is the total volume of solid
material. $\mathcal{P}$ denotes the porosity.
Should one of the materials within the grains reach its
sublimation temperature, it is replaced by vacuum and the optical
properties are adjusted accordingly in the model. In most cases
however, the disk is too cold for any material to sublimate.
Fig.\,\ref{fig:opacity} shows the mass opacity (absorption cross
section per unit mass) calculated using Eq.\,4 of \citet{Draine06} for
several prototypical dust compositions, with the size
distributions given in Tab.\,\ref{tab:dustmodels}.
%
%
%________________________________________________________________
\section{Modeling the Dust Continuum Emission}
\label{sec:dust}
\begin{figure}[h!btp]
  \includegraphics[angle=0,width=\columnwidth,origin=bl]{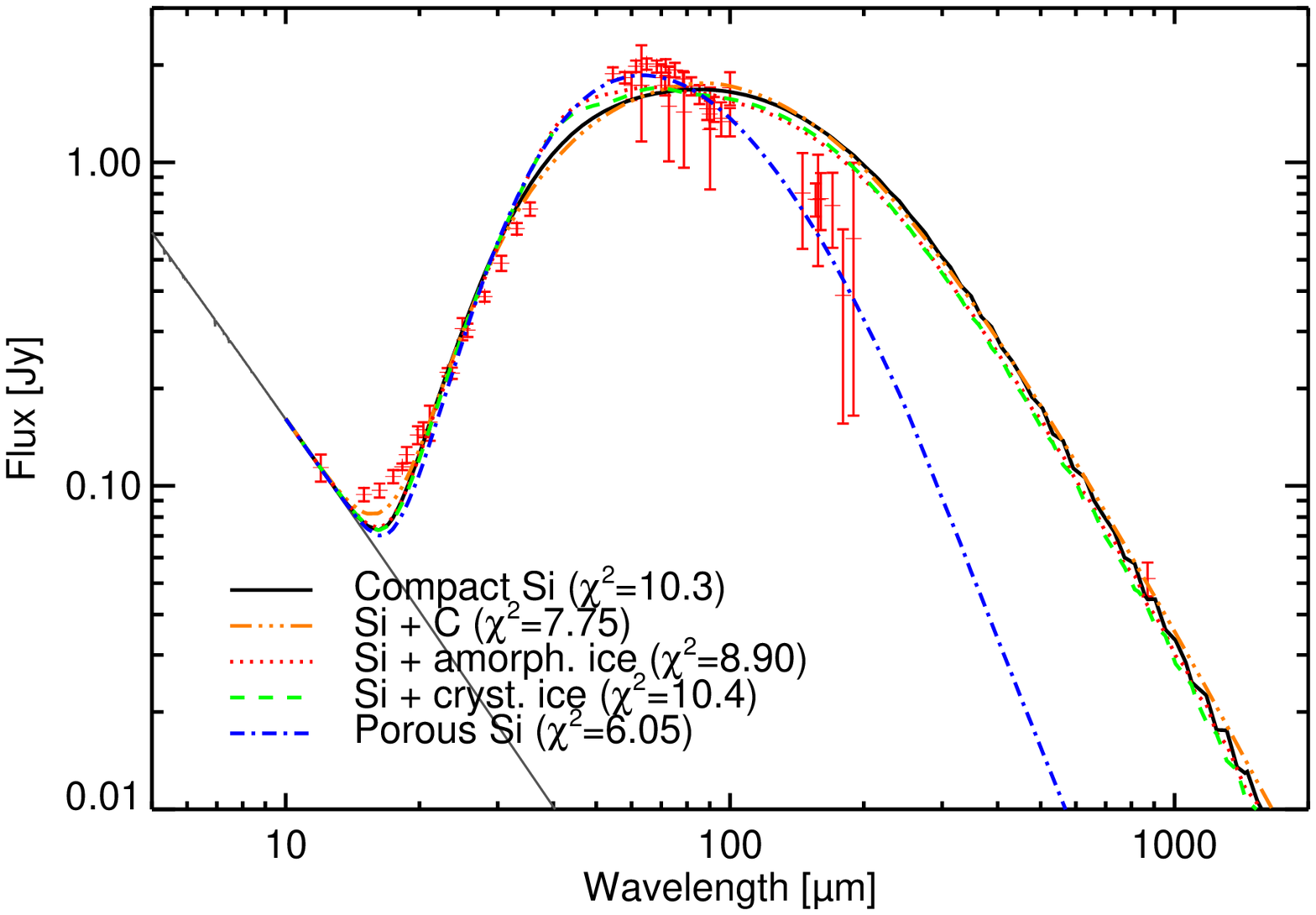}
    \includegraphics[angle=0,width=\columnwidth,origin=bl]{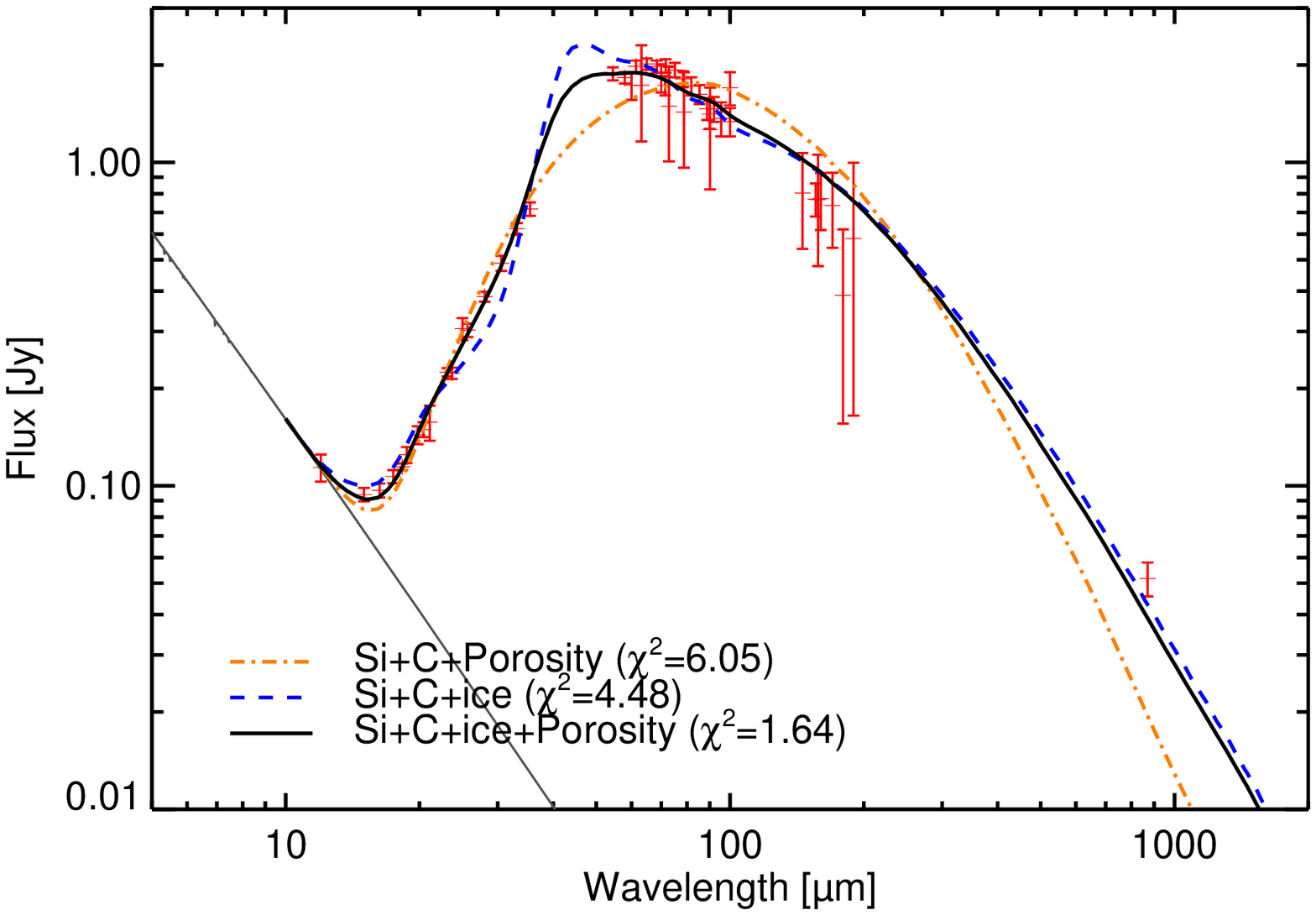}
    \caption{Best fit models (Table\,\ref{tab:dustmodels}) for several
      1- or 2-material mixtures ({\bf top}) and 3- to 4-material mixtures
      ({\bf bottom}). Red crosses: photometric data
      (Fig.\,\ref{fig:ObservedSED}).  Solid grey line: synthetic
      stellar spectrum.}
    \label{fig:SEDs}
\end{figure}

\begin{figure*}[hbtp] \label{fig:P_vs_size}
%\centering
\hspace*{-0.7cm}
\hbox to \textwidth{
\parbox{0.33\textwidth}{\includegraphics[angle=0,width=0.35\textwidth,origin=bl]{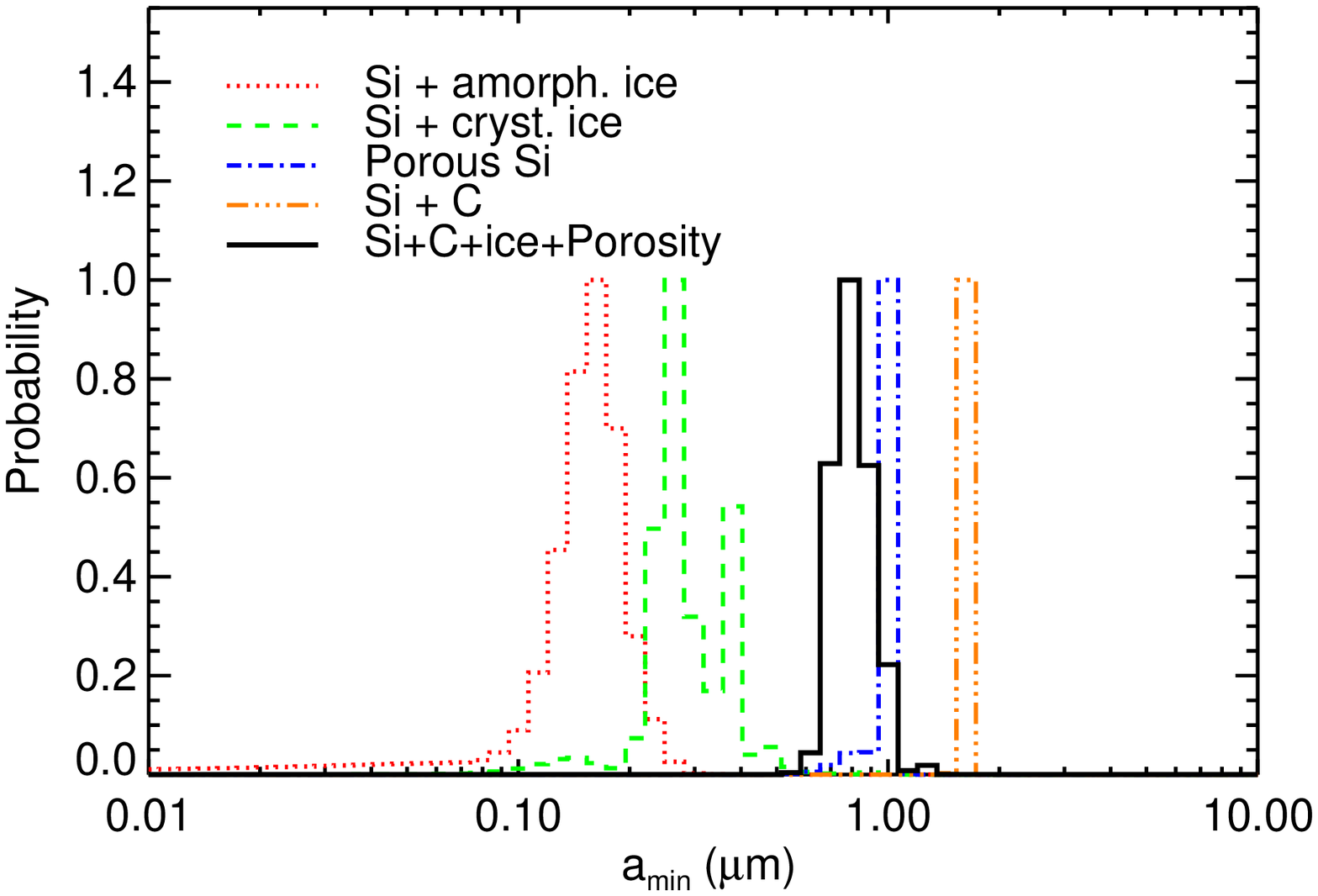}}
\parbox{0.33\textwidth}{\includegraphics[angle=0,width=0.35\textwidth,origin=bl]{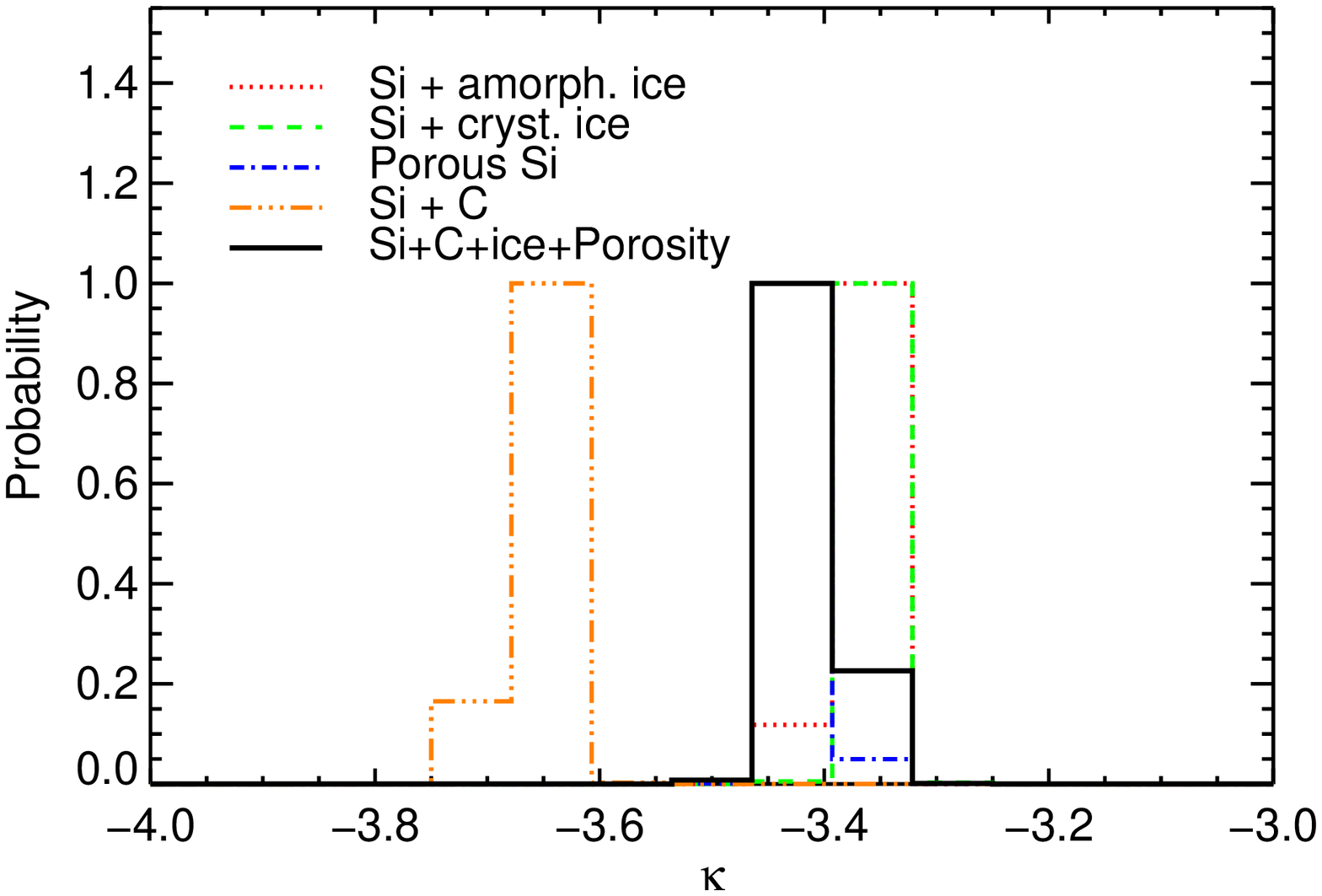}}
\parbox{0.33\textwidth}{\includegraphics[angle=0,width=0.35\textwidth,origin=bl]{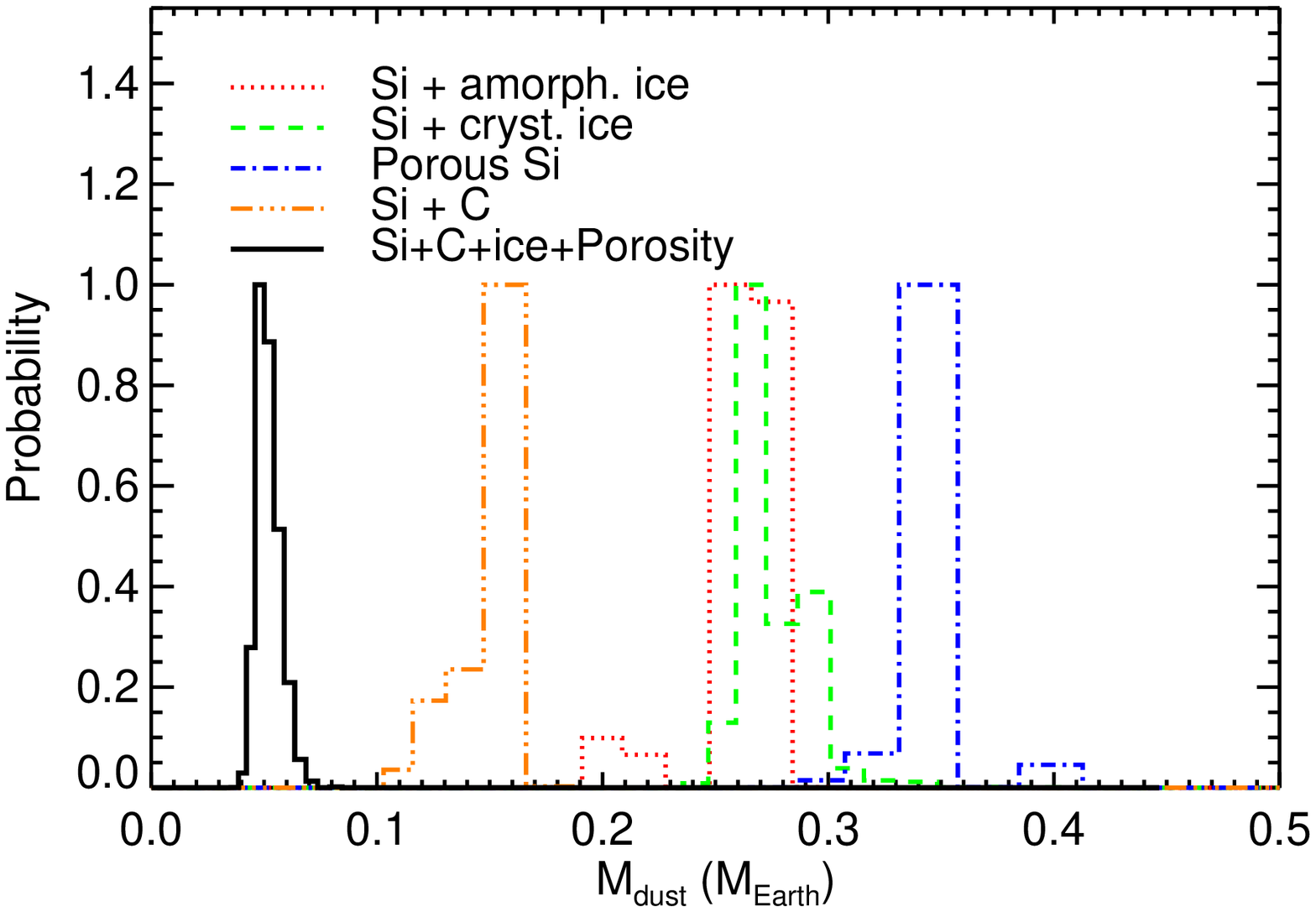}}
}
\caption{Bayesian probabilities for the models in Table\,\ref{tab:dustmodels}: minimum grain size, size distribution power law index, and total dust mass.}
\end{figure*}

We use the \gra\ code \citep{Augereau99a} to compute a grid of models
with a goal of reproducing the \hd\ Spectral Energy Distribution.  
\gra\ is a SED, image and interferometric data
fitter, specially designed to efficiently model optically thin disks
with parametric grain size and radial distributions, or
distributions from dynamical simulations.  The code accounts for both
the scattered light and continuum emission of dust grains
in thermal equilibrium with a star.
%  Each grain, characterized by its radius $a$, is assumed homogeneous
%and its thermal and light scattering properties are supposed to be
%well described by the Mie Theory valid for hard spheres.\

In the present case, the spatial dust distribution is constrained by
the resolved HST images (Sec.\,\ref{sec:RadialProfile},
Fig.\,\ref{fig:SDP}). The surface density profile derived
from inverting the observed surface brightness was imposed in the
model. Our study of the disk continuum emission focuses on
the properties of the dust particles, namely their size distribution
and composition. To that end, we computed large grids of models
reflecting various possible dust properties, starting from the
simplest composition (astronomical silicates) and progressively
increasing the model complexity until obtaining a good match to the
SED. For each model, the dust mass $M\dma{dust}$ and the quality of
the fitting are obtained from linear least-square fit between the
observed and the synthetic SEDs.  The $\chi^2$ listed in this
section are obtained using the 48 data points presented in
Tables\,\ref{tab:hd181327_phot_results},
\ref{tab:hd181327_spec_results} and \ref{tab:photometry}, and with the
numbers of degrees of freedom listed in
Table\,\ref{tab:dustmodels}. We subsequently use a statistical
(Bayesian) inference method to derive the probability distribution of
each parameter (see Appendix \ref{sec:statistics}).

In total we consider $\sim300$ compositions, with $77\times50$ grain
size distributions for each, leading to $\sim1,150,000$ models. All
the parameters of the grain models are summarized in
Table\,\ref{tab:graterparameters}.

%In summary, the dust model has 3 free parameters for the surface
%density and size distribution ($M_{\mathrm{dust}}$, $a_{\mathrm{min}}$
%and $\kappa$), and up to 2 additional free parameters (ice fraction
%and porosity $\mathcal{P}$) for the most advanced grain model.

\begin{table*}[btp]
%\begin{center}
  \caption{Best-fitting ({\it Best}, i.e. lowest $\chi\dma{r}^2$) and most
    probable ({\it Proba}, i.e. highest probability) parameters
    assuming different grain compositions (see text for the definition
    of the parameters). The most probable parameters are given with
    3-$\sigma$ uncertainties. The volume fractions marked with the
    $\dag$ symbol are imposed knowing the other free volume fractions.
    %
    %Volume fraction are defined such that
    %$v\dma{Si}+v\dma{C}+v\dma{ice} = 1$ and
    %${\mathcal{P}=V\dma{vac}/(V\dma{mat}+V\dma{vac})}$.  For each
    %composition, the first line indicates the expected value of the
    %parameters derived from bayesian analysis with , while the second
    %line corresponds to the best-fit parameters as extracted from the
    %grid \ie\ derived from $\chi\uma{2}$ minimization.
}
\label{tab:dustmodels}
\begin{tabular}{crccccccccc}
\hline
Composition & & $v\dma{Si}$ & $v\dma{C}$ &$v\dma{ice}$& $\mathcal{P}$& $a\dma{min}$(\um)& $\kappa$ &$M\dma{dust}$(M$\dma{\oplus}$)  & $\chi\uma{2}\dma{r}$ \\
\hline\hline \multicolumn{2}{c}{{\sc 1-material} (dof = 45)} & \\ \hline
\multirow{2}*{Si} & {\it Proba} : &{$ 1.00$}&--&-- & -- &  0.99$\pm$0.13 & -3.43$\pm$0.04&0.35$\pm$0.03& \\
& {\it Best} : &{$ 1.00$}&--&-- & --& {$ 1.00$} & {-3.42}&{0.35} & {$ 10.3$}\\
\vspace*{-0.2cm}\\
\multirow{2}*{C} & {\it Proba} : &--&1.00 & -- & --& 1.62$\pm$0.03 & -3.65$\pm$0.07 & 0.15$\pm$0.03 & \\
& {\it Best} : &--& {$ 1.00$}&-- & --& {$1.62$} & {$-3.64$}&{$0.15$} & {$ 7.75$}\\
\hline\hline \multicolumn{2}{c}{{\sc 2-material} (dof = 44)}  & \\ \hline
Si+C $^{(1)}$ & {\it Proba} :& 0.007$\dag$ & 0.99$\pm$0.04 &-- &  --&  1.62$\pm$0.03 & -3.65$\pm$0.07&0.15$\pm$0.03&  \\
\vspace*{-0.2cm}\\
\multirow{2}*{Si+am.ice} & {\it Proba} : & 0.59$\dag$& --& 0.41$\pm$0.05 &  --& 0.15$\pm$0.13 & -3.36$\pm$0.07&0.26$\pm$0.05&  \\
& {\it Best} : & {$0.60$}&--&{$0.40$}& -- & {$0.16$} &-3.35&{$0.26$}&  {$ 8.90$}\\
\vspace*{-0.2cm}\\
\multirow{2}*{Si+cr.ice} & {\it Proba} : & 0.67$\dag$& --& 0.33$\pm$0.12 &  --& 0.29$\pm$0.24 & -3.35$\pm$0.02&0.27$\pm$0.04&  \\
& {\it Best} : &{$0.65$} &--&{$0.35$}& -- & {$0.26$} &-3.35&{$0.27$}& {$ 10.4$}\\
\vspace*{-0.2cm}\\
\multirow{2}*{Si+$\mathcal{P}$} & {\it Proba} : &  0.94$\dag$& --&  -- & 0.057$\pm$0.062& 0.98$\pm$0.20 & -3.42$\pm$0.05& 0.35$\pm$0.04 &  \\
& {\it Best} : &{$ 0.025$} &--&--& {$0.975$} & {$0.010$} &-3.28 & 0.039 & {$ 11.9$}\\
\hline\hline \multicolumn{2}{c}{{\sc 3-material} (dof = 44)}  & \\ \hline
\multirow{2}*{Si+C+$\mathcal{P}$} & {\it Proba} : & 0.33$\dag$ & 0.67$\dag$ & -- &  0.95$\pm$6$\times10^{-5}$ & 9.45$\pm$3.20 & -3.13$\pm$0.09 & 0.042$\pm$0.001 & \\
	& {\it Best} : &{$0.33\dag$}&{$0.67\dag$}&-- & {$0.95$} & {$ 10.0$} & -3.14&{$0.042$}&  {$ 6.05$}\\
\vspace*{-0.2cm}\\
\multirow{2}*{Si+C+ice}& {\it Proba} : & {$0.07\dag$}&{$0.13\dag$}& 0.700$\pm$0.002 & -- &  0.78$\pm$0.04 & -3.50$\pm$0.03 &  0.11$\pm$0.01 & \\
	& {\it Best} : &{$0.07$\dag$$}&{$0.13\dag$}&{$0.80$} & -- & {$0.55$} & -3.42&{$0.11$}&  {$ 4.48$}\\
\hline\hline \multicolumn{2}{c}{{\sc 4-material} (dof = 43)}  & \\ \hline
\multirow{2}*{Si+C+ice+$\mathcal{P}$} & {\it Proba} : & 0.11$\dag$ & 0.22$\dag$ & 0.67$\pm$0.07  & 0.63$\pm$0.21& 0.81$\pm$0.31 & -3.41$\pm$0.09&0.051$\pm$0.016&  \\
& {\it Best} : &{$0.12\dag$}&{$0.23\dag$}&{$0.65$}& {$0.65$} & {$0.89$}&-3.42&0.048 & 1.64\\
\hline
\end{tabular}
%\end{center}
{\sc Notes --} ``dof'' means degrees of freedom (see Appendix\,\ref{sec:statistics}). $^{(1)}$ The best fit
for the Si+C mixture is obtained for silicate-free grains. The best-fitting 
parameters for that solution are given in the {\sc 1-material} part of
the table.
\end{table*}

%===================================================
\subsection{First Steps Toward a Complex Dust Model}
\label{sec:dustpreliminar}
%To better appreciate the need for a complex dust model, we start with
%fitting the \hd\ SED with a single- or 2-material grain model. 
We start by fitting the \hd\ SED with a single- or two-material grain model, which show the need for a more complex dust model.
% For that preliminary step, we simply use the nominal surface density
%profile with {\gHG~=~0.30} as described in Fig.\,\ref{fig:SDP}.
We consider several mixtures of the materials presented in
Sec.\,\ref{sec:dustproperties} to assess the broad properties of the
grains.
%pure silicate, pure organic refractories grains or 2-component
%mixtures (silicate/carbon, silicate/ice, silicate/vacuum). 
The model then includes 3 or 4 free parameters: (1) the minimum grain
size $a\dma{min}$; (2) the slope of the size distribution $\kappa$;
(3) the total mass of the disk $M\dma{dust}$; and in the case of
2-component grains: (4) the volume fraction of either carbon
($v\dma{C}$) or ice ($v\dma{ice}$), or the porosity ($\mathcal{P}$).
The detailed results of the investigation are summarized in
Table\,\ref{tab:dustmodels}, the best-fitting SEDs are displayed in Fig.\,\ref{fig:SEDs} (top panel) and probability curves are shown in Fig.\,9.

\subsubsection{Silicates or Organic Refractories?}
\label{sec:si+acar}
The most basic grain model to consider is the case of compact spheres
made of pure astronomical silicates (astroSi), a configuration that is
often used by default in many modeling approaches. We let
$a\dma{min}$, $\kappa$ and $M\dma{dust}$ vary and we compute a grid of
$\chi\uma{2}$ on which we apply a Bayesian analysis.  We find that
pure astronomical silicate grains would have
${a\dma{min}=0.99\pm0.13\,\mu}$m and ${\kappa=- 3.43\pm0.04}$, leading
to a total dust mass ${M\dma{dust} = 0.35\pm0.03 M\dma{\oplus}}$ for
grains smaller than 1\,mm (Tab.\,\ref{tab:dustmodels}).  These values
are consistent with the ones typically found for debris disks, but
they actually lead to very poor fits,
%The global shape of the SED is poorly
%explained assuming such a material 
as illustrated in Figure\,\ref{fig:SEDs} (top panel) where the model
with the smallest $\chi\uma{2}$ ($\chi^2\dma{r}$ = 10.3) is displayed.
Although the model is in fairly good agreement with the data over the full spectral range
and agrees well with the sub-mm measurement, it fails to reproduce
the shape of the emission peak at $\sim50-100$\,\um\, and the mid-IR
fluxes, shifting the dominant part of the disk to colder temperatures.
It overestimates by 30-40\% the excess at far-IR wavelengths ($\lambda
\sim 100-160$\,\um), in particular the Herschel/PACS measurements.

%
% The total
%mass of the dust disk would be $0.35 M_{\oplus}$ (for grains up to
%1 mm) with a blowout size $R\dma{blowout} = 0.89\,$\um, in good
% agreement with the minimum grain size.  
%
%The maximum optical depth
%$\tau\dma{perp}$ = 0.032 is fully consistent with the assumption of an
%optically thin disk and the typical collision timescale of $\sim 7400$
%years at the peak density position confirms that we are dealing with a
%collision-dominated debris disk. (see discussion in
%Sec.\,\ref{discu:timescale}).
%% ASTROSI + ACAR
We take a further step and consider mixtures of amorphous
silicates and carbonaceous materials.  We ran a Bayesian analysis and
find that the actual best composition corresponds to pure carbon.
Qualitatively, the effect on the SED is not significant (see
Fig.~\ref{fig:SEDs}, top panel), even if it leads to a better minimum
chi-square ($\chi^2\dma{r} = 7.75$). This model gets closer to the
measured fluxes at the peak emission ($\lambda \sim50-100$ \um), but
does not provide an overall good fit. The addition of carbonaceous
material slightly shifts the best fit parameters of the grain size
distribution (Tab.\,\ref{tab:dustmodels}), resulting in slightly
larger $a\dma{min}$ values and slightly steeper size distributions
%grains at the lower limit of the size distribution %($a\dma{min} =
%1.62\pm0.03\,\mu$m), but in larger amount. %($\kappa = -
%3.65\pm0.07$).
%The total mass of the dust disk, ${M\dma{dust} = 0.15\pm0.03
%M\dma{\oplus}}$, logically reduces as the density of carbon is
%smaller than the one of silicates.

\subsubsection{Icy Grains?}
\label{sec:astrosi+ice}
%%%%%%%%%%%%%%%%%%%%%%%%%%%%%%%%%%%%%% AMORPHOUS ICE %%%%%%%%%%%%%%%%%%%%%%%%%%%%%%%%%%%%%%
\citet{Chen08} suggested that HD 181327 may contain a significant
fraction of water in the form of crystalline ice, producing a broad
peak in the combined Spitzer/IRS (5-35\,\um) and MIPS-SED (60-77\,\um)
spectrum.
To assess the presence of water ice in the debris disk of \hd, we
model the disk with grains made of a mixture of astronomical silicates
and water ice, either in the form of amorphous ice or crystalline ice.
%
%To compute the likelihood function of the ice volume fraction we
%integrate the probability function over $\kappa$ and $a\dma{min}$
%following the methodology of Appendix\,\ref{sec:statistics}.  

The results of the statistical analysis are documented in
Table~\ref{tab:dustmodels}.
The quality of the fit is only slightly improved compared to
the pure silicate model: the best models have $\chi^2\dma{r} = 8.90$ for
amorphous ice, $\chi^2\dma{r} = 10.4$ for crystalline ice.
In fact, the addition of water ice, crystalline or
amorphous, does not modify the general shape of the synthetic SED (see
Fig.\,\ref{fig:SEDs}, top panel).
%
%of the results leads to strong constraints on the properties of the
%grains: if the water ice is amorphous, then it represents $\sim$40\%
%of the volume of the dust grains ($v\dma{ice} = 0.41\pm0.05)$, which
%are particularly small.
%
%Crystalline water ice would represent a similar, possibly slightly
%smaller, volume fraction of the grains material ($v\dma{ice} =
%0.32\pm0.12)$ and would increase the minimum grain size.  In both
%cases, we find a similar slope and, since most of the mass is
%contained in the largest grains, this results in a similar total disk
%mass. 
The smallest $\chi\uma{2}$ calculated with amorphous ice is slightly
smaller than the one of crystalline ice, but in both cases the values
are too high to accept these models.

%Such a population of grains would be surprisingly small with $a\dma{min} =
%0.08$\,\um, whereas they should have been ejected from the system by
%the radiation pressure ($R\dma{blowout} = 1.28$\,\um).  The power-law
%exponent remains close to the theoretical value with $\kappa = -3.43$.

%%%%%%%%%%%%%%%%%%%%%%%%%%%%%%%%%%%%%% CRYSTALLINE ICE %%%%%%%%%%%%%%%%%%%%%%%%%%%%%%%%%%%%%%

%We see that the presence of amorphous ices improves the fit for an ice
%fraction of 0.5. We run 19 grids of models for crystalline ice
%fractions ranging from 0.0 to 0.95.  We find a best fit model for a
%crystalline ice fraction of 0.4, for which the ${\chi}^2$ is 502.0
%($\chi^2\dma{r}$ = 11.2).  The total mass of the dust disk and the
%slope of the grain size distribution are close to the ones found for
%amorphous ice ($M\dma{dust}$ = 0.21 $M_{\oplus}$, $\kappa = -3.43$),
%and so is the blowout size ($R\dma{blowout} = 1.13$\,\um).  The
%minimum grain size however is four times larger ($a\dma{min} =
%0.30$\,\um) than the one of amorphous ice.  Hence despite a $\chi^2$
%value not as good as the one found for amorphous ice, the crystalline
%grains case is easier to explain thanks to the best agreement between
%$a\dma{min}$ and $R\dma{blowout}$.

\subsubsection{Porous Grains?}
\label{sec:porous}
As a final step in our preliminary study of the dust composition, we
consider a grid of models consisting of silicate grains with
different porosities, from compact silicates to extremely fluffy
silicate grains ($\mathcal{P}\dma{max} = 99.375\%$).  The marginal probability distribution for the porosity $\mathcal{P}$ reveals a bimodal probability distribution: the most probable models
are the ones with either no porosity (discussed in
Sec.\,\ref{sec:si+acar}), or very high porosity ($\mathcal{P} =
97.5\%$). Both solutions are presented in Tab.\,\ref{tab:dustmodels}.
% There is a secondary peak in the probability distribution for
%$\mathcal{P} = 97.5\%$.
A visual inspection of Figure~\ref{fig:SEDs} resolves the
ambiguity.  The compact grain model better reproduces the minimum and
maximum wavelength ends of the SED, and in particular 
fits the sub-mm point, but it fails at reproducing the intermediate
wavelength range around the emission peak.  On the other hand, the
highly porous grain model is orders of magnitude below the observed
value at 870\,\um, but it is within the error bars in the other
regions of the SED, especially at \her\ wavelengths.
% that reflect more significantly the dust population. 
In sum, none of these models lead to overall good fits to the
whole SED.
\subsection{Construction of an Advanced Grain Model}
\label{sec:ism}
From what precedes, we conclude that none of the 2-component mixtures
considered are sufficient to provide a satisfactory fit to the SED. This is
essentially a consequence of the disk surface density being known \textit{a priori} in
this model; it cannot be adjusted to compensate for the emission
properties of the dust grains. We note, however, that the addition of
one component to silicates, should it be carbonaceous material,
amorphous ice, or porosity, provides some slight improvements. This
suggests that their combined effect could produce a more realistic
model.

% Figure : Bayesian compo
\begin{figure*}[h!tbp]
%\centering
\hspace*{-0.7cm}
\hbox to \textwidth{
\parbox{0.33\textwidth}{\includegraphics[angle=0,width=0.35\textwidth,origin=bl]{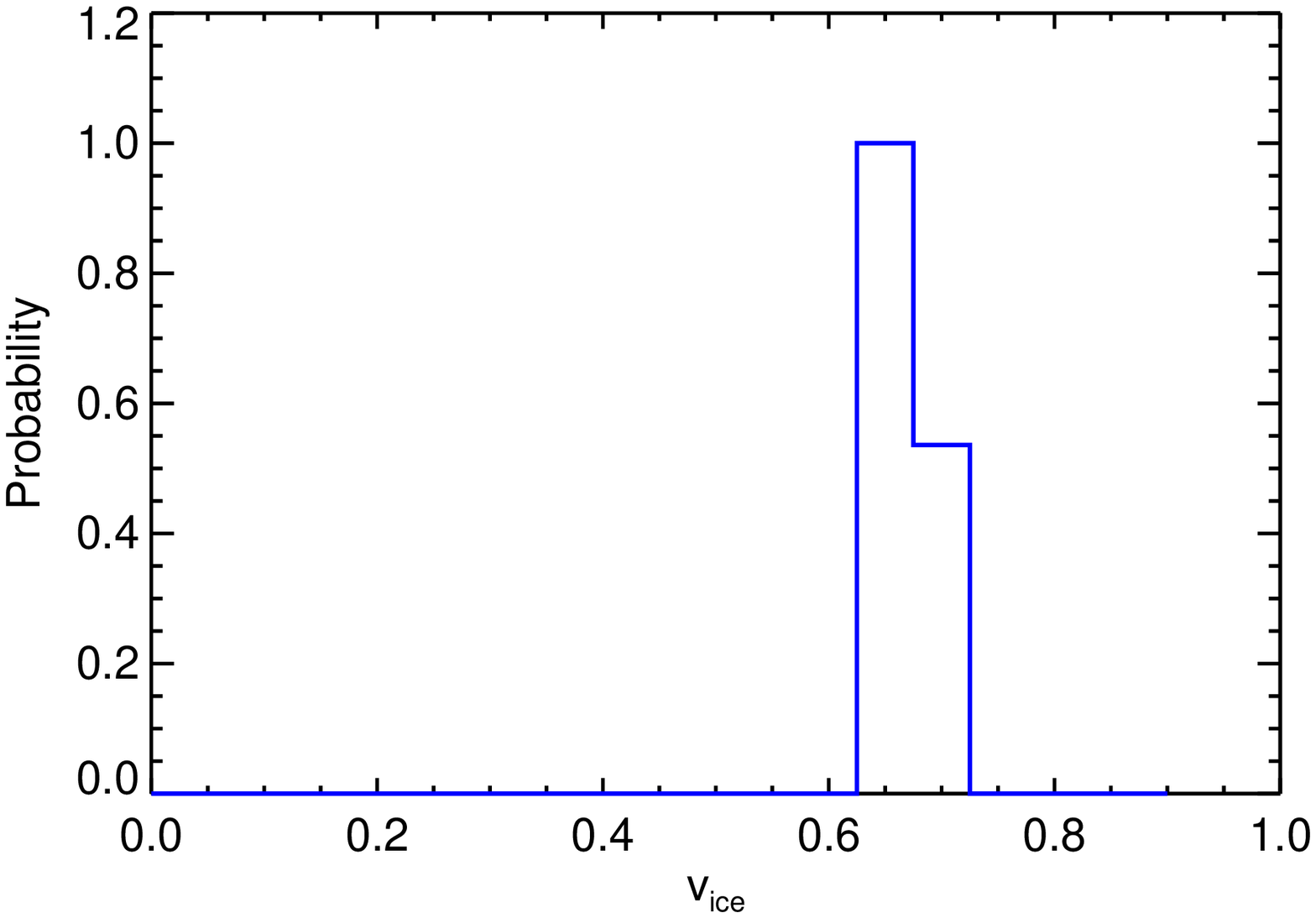}}
\parbox{0.33\textwidth}{\includegraphics[angle=0,width=0.35\textwidth,origin=bl]{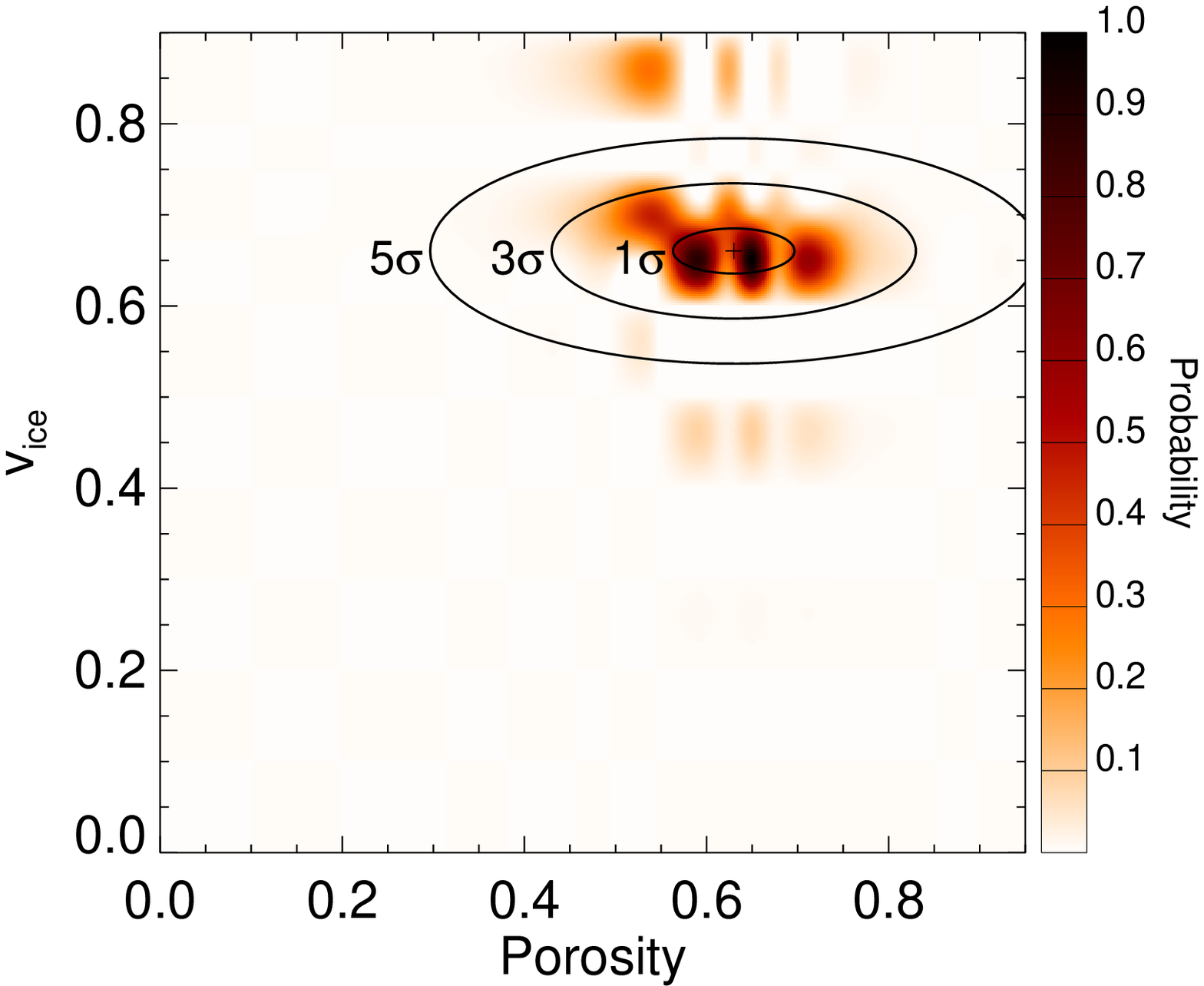}}
\parbox{0.33\textwidth}{\includegraphics[angle=0,width=0.35\textwidth,origin=bl]{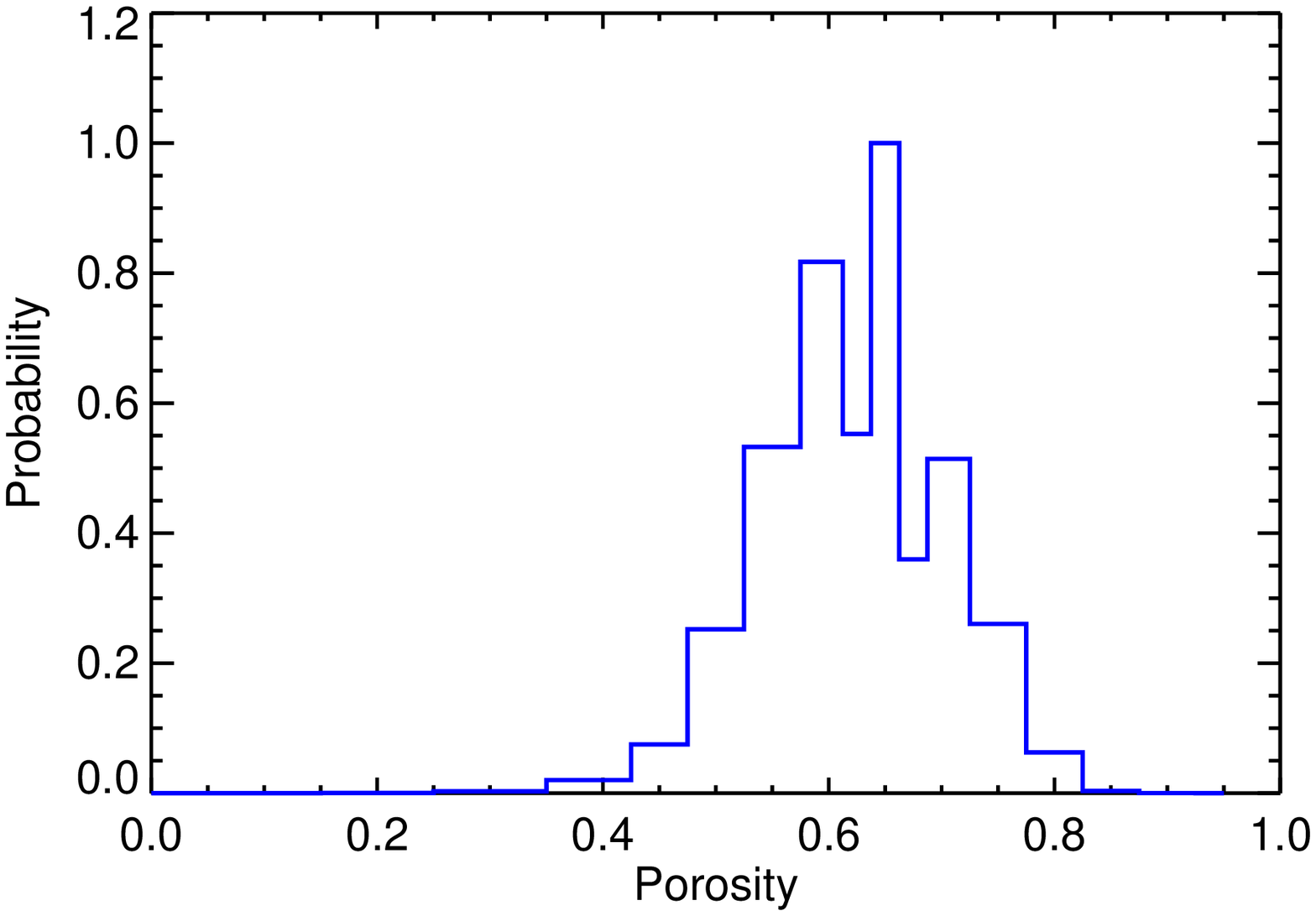}}
}
\caption{Bayesian probabilities for the grain composition in the case of 4-material dust mixtures.}
\label{fig:P_vs_compo} 
\end{figure*}

We evaluate the improvement provided by a 3-component mixture on the
fitting of the SED. We showed in Sec.\,\ref{sec:si+acar} that it is not clear
how to disentangle the respective contribution of silicates and
organic materials by looking at the SED.  Instead of including an
additional free parameter, we consider grains (approximately) in fixed Si/C ratio described in
Sec.\,\ref{sec:dustproperties} ($v\dma{C}/v\dma{Si} = 2$), and we restrict
the discussion to the ice fraction and the porosity of the grains. We
thus assume that grains are made of silicate and carbonaceous material, which
are either icy ($v\dma{ice}$ parameter) or porous ($\mathcal{P}$
parameter). The results are reported in
Tab.\,\ref{tab:dustmodels}. For porous ice-free grains, the best model
is found for a porosity $\mathcal{P}$ = 95\% and leads to some
improvement to the fit with $\chi\uma{2}\dma{r}=6.05$. If we now
consider only compact icy grains, we can achieve a very reasonable
$\chi\uma{2}\dma{r} = 4.48$ with comparison to previous models,
assuming an amorphous ice fraction $v\dma{ice} = 0.7$. Overall, both
the addition of either water ice or vacuum to silicate + carbonaceous
grains do contribute to improving the fit to the SED of \hd. This can
can be seen in Figure~\ref{fig:SEDs} (bottom panel). Nevertheless, the
fits are still unsatisfactory ($\chi\dma{r}^2$ much larger than 1),
and justify an even more advanced dust model.

We next consider a 4-component grain model inspired by the ``cold
coagulation dust model'' by \citet{LiLu03a,LiLu03b}, which has been
successfully employed for modeling debris disks \citep{LiGreenb,
  Augereau99a, LiLu03a,LiLu03b,kohler2008}. We keep the Si/C volume
ratio fixed ($v\dma{Si}$:$v\dma{C}$=$1$:$2$) and let both the ice
fraction and the porosity vary, in addition to the other usual free
parameters in this study, namely the minimum grain size $a\dma{min}$,
the slope of the size distribution $\kappa$, and the total mass of the
disk $M\dma{dust}$. This leads to a five-free-parameter disk model to fit
the 48 data points in the SED. We compute three grids of models corresponding to
the three surface density profiles shown in Fig.\,\ref{fig:SDP} for
\gHG~=~0.30 (i.e. the nominal and $\pm 1\sigma$ profiles).  The
Bayesian analysis is then applied to the 3 grids separately and the
probabilities for each point of the parameter space are calculated as
the sum of the probabilities for the 3 profiles.  The detailed results
are documented in Tab. \ref{tab:dustmodels}.

Figure\,\ref{fig:P_vs_compo} presents the results of the
Bayesian analysis on the composition of the dust grains, with a 2D-map
showing the probability as a function of $\mathcal{P}$ and
$v\dma{ice}$ for the grid models, after projection of the probabilities over the other
parameters.  It clearly illustrates the need for a
composition within a limited range of possible values.  In
particular, the fraction of H$_2$O-dominated ices is well defined: the
probability peak is obtained for an ice fraction $v\dma{ice} =
0.67\pm0.07$.  The porosity is not so strongly constrained but we
find a peak of probability for $\mathcal{P}=0.63\pm0.21$. 
The best fit model in terms of $\chi\uma{2}$ is found for grains made
of 23\% carbonaceous materials, 12\% silicates, 65\% ice,
and with a porosity $\mathcal{P} = 65\%$. As can be seen in the bottom
panel of Figure~\ref{fig:SEDs}, for such grains, the synthetic SED
very reliably describes the observed SED from the mid-IR to the sub-mm, with ${\chi\uma{2}\dma{r} = 1.64}$, a value significantly
better than those obtained for simpler grain models.
The grain size distribution derived from the statistical analysis
has $a\dma{min} = 0.81\pm0.31$ \um\ and $\kappa = -3.41\pm0.09$, a
power law index very consistent with that predicted for theoretical collisional
cascades ($\kappa = -3.5$). The mass in grains smaller than 1 mm,
$M\dma{dust} = 0.051\pm0.016\,M\dma{\oplus}$, may look surprisingly
small at first glance compared to the other models, but it results
from the significant fractions of ice and vacuum in the grains,
implying larger mass opacities (Fig.\,\ref{fig:opacity}). The
temperature of the grains at the surface density peak position (89.5
AU) ranges from 86 K for the smallest grains to 40 K for $\sim$ 8 mm-sized
grains.

To summarize, our approach, based on a fixed surface density
profile constrained by resolved imagery of the disk, clearly
highlights the need to consider the presence of both
ice and porosity to find an overall good fit to the SED. Icy grains
are naturally expected at such a large distance from a solar-type star
and the fluffiness of the grains is likely a by-product of their formation process. 
We applied a similar cold-coagulation dust model as \citet{LiLu03a} did fitting the very young \object{HD\,141569A} debris disk, 
except that we used no \textit{a priori} assumption on the ice fraction,
but with arguing the grain composition must reflect the primordial abundances of condensable elements, they chose 
$v\dma{ice} \approx 30\%$ %($m\dma{ice}/(m\dma{Si}+m\dma{C}) \approx 0.7$) 
for cold regions 
and they find a best-fitting porosity $\mathcal{P} \approx 0.8$. % very similar to one we infer.
%$\sim 3.5 M\dma{\oplus}$ with a porosity $\mathcal{P} = 0.90$ 
%(from 1 \um\ to 1 cm with $\kappa$ = -3.3) for the 141569A debris disk. 
Hence we propose a dust model that is especially ice-rich with a degree of porosity that is not unexpected.  
Further comparisons with the literature and 
implications on the nature of the dust are discussed in more 
details in Sec.\,\ref{sec:discu}. 
%________________________________________________________________
\section{Limits on the Gas Content}
\label{sec:gas}
\label{sec:gas}
\begin{figure}[h!btp]
\begin{center}
 \hspace*{-0.4cm} \includegraphics[angle=0,width=1.05\columnwidth,origin=bl]{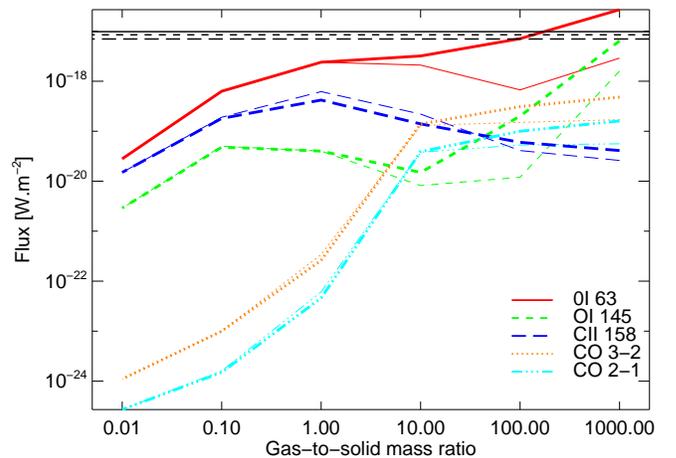}
  \caption{Expected gas line fluxes as a function of gas-to-solid mass ratio for either a low-PAH abundance (thin lines) or a high-PAH abundance (thick lines). The horizontal lines denote the PACS 3$\sigma$ upper-limits and do not include the 30\% flux calibration uncertainty. The gas-to-solid mass ratios assume a
    solid mass of 0.164 M$_{\oplus}$.}
    \label{fig:prodimo}
  \end{center}
\end{figure}

{At 12 Myr-old, \hd\ offers a chance to assess the amount of gas
in a young debris disk, at a time when, according to planet formation theories, 
the accretion of gas by giant planets has already occurred 
while the formation of rocky planets might be ongoing. 

We use the upper limits on the oxygen and ionized carbon fine-structure lines  
($\mathrm{[O~{\sc I}]}$ at 63 and 145 $\mu$m, and $\mathrm{[C~{\sc
II}]}$ at 158 $\mu$m, see
Tab.\,\ref{tab:spectro}) to constrain the amount of gas present in the disk.
We solve the photochemistry and energy balance in the disk using the photochemical code 
ProDiMo\footnote{\citet{Woitke2009A&A...501..383W} provide a
detailed description of the code, with subsequent improvements
described in \cite{Kamp2010A&A...510A..18K},
\cite{Thi2011MNRAS.412..711T}, and \cite{Woitke2011arXiv1103.5309W}.}.
The abundance of 71 species is computed in steady-state, under the influence of adsorption, desorption (thermal, photo and cosmic-ray induced), and molecular hydrogen formation on grain surfaces.

The gas density profile is supposed to match the one of the dust, as presented in Sec.\,\ref{sec:RadialProfile}.
The total mass of \textit{solid} (defined as the mass in grains from 0.81\,$\mu$m to 8\,mm) is constrained by the 
best dust model of Sec.\,\ref{sec:ism} ($M\dma{solid}$ = 4.9~$\times$~10$^{-7}$ M$_\odot$ =~0.164 M$_{\oplus}$)
and we let the total atomic+molecular gas mass
$M_{\mathrm{gas}}$
vary from a gas-poor disk ($M\dma{gas}/M\dma{solid}$=10$^{-2}$)
to a gas-rich disk ($M\dma{gas}/M\dma{solid}$=10$^{3}$).
The parameters of the model are summarized in Tab.\,\ref{tab:prodimo}.
One of the main parameters controlling the
gas energy balance is the amount of PAHs ($f_{\mathrm{PAH}}$), which remains unconstrained in the disk of \object{HD~181327}
because their abundance is too small to produce detectable PAH emission features in the IR. 
We adopted two "extreme" values for $f_{\mathrm{PAH}}$: a
low PAH abundance ($f_{\mathrm{PAH}}$=10$^{-5}$) and a high PAH
abundance ($f_{\mathrm{PAH}}$=0.1)\footnote{The high PAH abundance is an
upper-limit to the amount of PAHs in HerbigAe disks, where it is found
to be of the order of 0.1-0.01 (e.g.,
\citealt{Meeus2010A&A...518L.124M}, $f_{\mathrm{PAH}}$=0.03 for
\object{HD 169142}).}.

The non-LTE line fluxes for the $\mathrm{[O~{\sc
    I}]}$, $\mathrm{[C~{\sc II}]}$, CO $J$=3 $\rightarrow$ 2, and CO
$J$=2~$\rightarrow$~1 transitions are given in Fig.\,\ref{fig:prodimo} and Appendix\,\ref{sec:tab_gas}.
Among all the lines, only the {CO~$J$=3~$\rightarrow$~2} and {$J$=2~$\rightarrow$~1}
line fluxes behave monotonically, increasing in flux
with increasing disk mass. The reasons for the apparent erratic
behaviour of the fine-structure lines come from the non-linear nature
of several physico-chemical phenomena. For example, the
$\mathrm{[C~{\sc II}]}$ flux depends on the abundance of C$^+$ and on
the excitation of the first energy level. The C$^+$ abundance first increases
with total gas mass until CO self-shielding against photodissociation
is efficient enough such that a significant fraction of C$^+$ is
converted into neutral C and CO. The presence of CO also changes the
heating and cooling balance, since the CO absorption of IR stellar
photons becomes the main source of heating. The change from a disk
with no CO self-shielding to a CO self-shielded disk manifests itself
as a jump in CO fluxes between a disk with gas-to-solid ratio of 1 to a
disk with gas-to-solid ratio of 10.  The abundance of PAHs has a weak
influence on the gas temperature because the small UV flux from a F5/F6
star without excess accretion luminosity cannot ionize PAHs.
On the other extreme, when the gas mass is very small (at gas-to-solid
mass ratio below 1), the PAH and molecular absorption heating are
nonexistent. Photoelectric heating by large grains dominates and
explains why the line fluxes are similar for the high and low PAH
abundance models.

Line fluxes from models with $f_{\mathrm{PAH}}$=10$^{-5}$ are
consistent with the PACS upper limits for all gas-to-solid mass ratios
(upper part of Table~\ref{tab:gas}).  
Only in the case
$f_{\mathrm{PAH}}$=0.1, the disk gas mass is constrained by [OI] at 63 \um\
to be below
$\sim$17 M$_{\oplus}$ ($M\dma{gas}/M\dma{solid}~\lesssim~10\uma{2}$, lower part of Tab.~\ref{tab:gas}).

Eventually, even if the integration of the $\mathrm{[O~{\sc I}]}$ 63 and 145 $\mu$m, and $\mathrm{[C~{\sc II}]}$
lines we performed was not sufficient to derive strong constraints on the gas-to-solid mass ratio in the disk, the combination of the GRaTeR 
and ProDiMo codes proved efficient to make predictions on the fluxes in the 
{CO~$J$=3~$\rightarrow$~2} and {$J$=2~$\rightarrow$~1} lines. These lines will be detectable for instance by ALMA, finally allowing to assess the gas content of the debris disk.

%%%% Table: Prodimo parameters
\begin{table}[h!btp]
%\begin{center}
  \caption{Setup of ProDiMo for the gas disk model}
\label{tab:prodimo}
\begin{tabular}{lcc}
  \hline
  Parameter & Value \\
  \hline
  Elemental abundances & ISM-like \\
UV field & 1 Habing field $\uma{(a)}$ \\
Cosmic-ray ionization rate $\zeta$ & 1.7~$\times$~10$^{-17}$ s$^{-1}$ $\uma{(b)}$ \\
Non-thermal line width $dv$ & 0.15 km.s$^{-1}$ \\
$f_{\mathrm{PAH}}$$\uma{(c)}$ & 10$^{-5}$ - 10$^{-1}$ \\
     \hline	
\end{tabular} \\
$\uma{(a)}$ \citet{2006PNAS..10312269D}, 
$\uma{(b)}$ Standard interstellar UV field,
$\uma{(c)}$ $f_{\mathrm{PAH}}$ scales with the gas mass and is defined with respect to an interstellar abundance of 3~$\times$~10$^{-7}$.
\end{table}
}

%________________________________________________________________
\section{Additional constraints on the disk}
\label{add_constraints}
\label{sec:discu}
\subsection{Scattered light: evidence for irregular aggregates}
\label{sec:scattering}
The surface density profile used in this study comes from a new
reduction of the scattered light images of \citet{Schneider06}
obtained with HST/NICMOS (see Sec.\,\ref{sec:RadialProfile}). This
ensures by definition that the shape of any synthetic scattered light
surface brightness profile in this study perfectly agrees with the
observed profile at $1.1\,$\um. However, the consistency is limited to
the shape, and the agreement in terms of absolute flux is not
guaranteed. This is because our fitting approach in
Sec.\,\ref{sec:dust} consists of adjusting the disk mass to get a best
fit to the SED, which is equivalent to scaling up or down the surface
density profile.
%in Sec.\,\ref{sec:dust} is limited to the longest wavelengths of the
%SED which correspond to the light thermally emitted by the dust
%particles.  
A reason for adopting this strategy comes from the fact that the
scattered light images do not directly provide the surface density
$\Sigma(r)$, but its product by the mean scattering cross section
$\sigma\dma{sca}$ at the observing wavelength ($1.1$\,\um). The
fitting process to the SED, on the other hand, yields an independent,
absolute scaling of the surface density profile, thus allowing one to obtain an
initial estimate of $\sigma\dma{sca}$.
%. It is thus possible to get a first estimate of $\sigma\dma{sca}$ by
%dividing the quantity $\sigma\dma{sca}\Sigma(r)$ obtained from the
%scattered light images (Sec.\,\ref{sec:RadialProfile}) and
%$\Sigma(r)$ obtained from the SED fitting (Sec.\,\ref{sec:dust}).
We find $\sigma\dma{sca} = 3.7\,$\um$^2$ in the NICMOS $1.1\,$\um\
band.

A second estimate of $\sigma\dma{sca}$ is obtained by calculating the
theoretical scattering cross section for compact spherical grains
given the grain properties (composition, size distribution) that were
inferred in Sec.\,\ref{sec:dust}:
%\begin{equation}
%\sigma\dma{sca}(\lambda) = \int\limits_{a = a\dma{min}}^{a\dma{max}}
%{\pi a^2 Q\dma{sca} (a, \lambda) \mathrm{d}n(a)}
%\end{equation}
${\sigma\dma{sca}(\lambda) = \int_{a = a\dma{min}}^{a\dma{max}} {\pi
a^2 Q\dma{sca} (a, \lambda) \mathrm{d}n(a)}}$ where $Q\dma{sca}$ is the
scattering efficiency at wavelength $\lambda$ and for a grain radius
$a$.  For the best model (in the Bayesian sense,
Tab.\,\ref{tab:dustmodels}) we find $\sigma\dma{sca} = 16.8\,$\um$^2$
at $1.1\,$\um, a factor of about 4.5 higher than the first
estimate. This implies a peak flux density of 3.14\,mJy.arcsec$^{-1}$
inconsistent with the 0.76 mJy.arcsec$^{-1}$ measured in the NICMOS
profile.
%To account for the measured surface brightness peak in the framework
%of the Mie theory, a mean scattering cross section $\sigma\dma{sca} =
%3.7 \mu m^2$ would be required, implying either smaller grains or
%smaller optical indexes.
%
In other words, this means that our model predicts an albedo ($\omega
= 59\%$ at $1.1\,$\um) that is a factor of about 4.5 larger than the
observed value. This suggests that smaller grains than those required
to fit the SED dominate the scattering behavior of the disk.
%This finding
%is consistent with the results of \citet{Schneider06} who similarly
%identified an inconsistency between different diagnostics for the size
%distribution.  They in particular show that the rather low scattering
%asymmetry factor $|g|$ of \hd\ requires grains that are far too small
%to be consistent with other disk properties (colors, SED, dynamical
%constraints). 
It is worth noting that such a conclusion is not specific to \hd, but is
also observed in other debris disks. Recent examples
include \object{HD\,207129} for which \citet{Krist2010} predict an
albedo (based on dust properties also obtained from a fit to the SED)
that is an order of magnitude larger than the value suggested by the
scattered light image. A similar inconsistency is noticed
for \object{HD\,92945} by \citet{gol11} who point out that different
observing diagnostics suggest different minimum grain size.

It is important to keep in mind that although our model suggests the
grains are very porous and hence likely to have an irregular, fractal
structure, we used hard spheres (Mie theory) to calculate both the
thermal emission and scattered light optical properties. Unlike the
thermal behavior, light scattering will likely strongly depend on the exact
shape of the dust aggregates, and on the size of their elemental
constituents, two features that are not accounted for by the Mie
theory \citep{vosh07}. Another piece of
evidence that Mie theory might be a limiting factor for a
self-consistent model of \hd\ comes from the observed scattering
asymmetry parameter $g\dma{HG}$ that is a measure of the direction of
the light scattered by a particle.
%scattering asymmetry as seen by the HST. The scattering asymmetry
%parameter, approximated by the Henyey-Greenstein phase function
%$g\dma{HG}$, is a measure of the direction of the light scattered by a
%particle. 
For our best model (Sec.\,\ref{sec:ism}) the Mie theory yields a mean scattering asymmetry
parameter \gHG$ = 0.95$ at $\lambda = 1.1\,$\um, much larger than
the \gHG$ = 0.30$ measured on the NICMOS image.  Only a population of
much smaller particles could produce such a low scattering asymmetry
parameter if we stick to the Mie formalism, as already noted by
\citet{Schneider06}. An analogous problem arises from the modeling of
the debris disk of \object{HD\,92945} \citep{gol11}.
%
%surrounding the T Tauri star IM Lupi by \citet{Pinte2008}, who
%oppositely predict too little anisotropy with respect to the scattered
%light image, due to the small size of their grains. 
Our micron-sized grains are highly anisotropic, essentially due to
their spherical shapes.  Incorporating vacuum homogeneously inside the
grains does not suffice to decrease \gHG\ and a more accurate
description of the shape of fluffy aggregates is eventually needed.
In fact, the scattering efficiencies and asymmetry parameters of
fluffy aggregates are known to be correctly described by an EMT-Mie
model (porous spheres) only in the Rayleigh regime, \ie\ when the size
of the inclusion (vacuum here) is small compared to the
wavelength \citep[see {e.g.}][]{vosh07}. This is not the case
for the relatively large grain sizes and porosity 
when the wavelength becomes small.

%A natural explanation to the excess of scattering we predict, is that
%the scattering properties of porous aggregates cannot be modelled
%correctly by the Mie theory.

%This shows that a simplistic spherical geometry is not fully suited
%to describe fluffy aggregates.  The adequacy of the Mie theory to
%describe the scattering properties of dust aggregates has been
%studied by several authors.  \citet{Shen2009} compared different
%approaches to model the light scattering properties of random
%aggregates, in particular they compare the EMT-Mie model calculated
%with the Bruggeman mixing rules and the more accurate Discrete Dipole
%Approximation.  For the range of grain sizes they consider, they
%conclude that the scattering phase function is essentially dependent
%on the typical size of the aggregates and not on their porosity, they
%also point that the EMT-Mie calculation tends to produce insufficient
%backscattering for short wavelengths: the phase function in the
%EMT-Mie approach is slightly underestimated in the backward
%direction, while it is seriously overestimated in the forward
%direction, resulting in overestimated values for \gHG.\

\subsection{A narrower parent belt?}
\label{sec:discusslongwav}
Unlike scattering, we expect the continuum thermal emission to be largely independent of the exact shape of the grains. 
Our best model of \hd's disk produces a sub-mm flux of 39\,mJy at
870\,\um, consistent within 2$\sigma$ with the LABOCA measurement
($52\pm6$\,mJy).  
The ATCA measurement at 3.2\,mm was not model-fitted, because of large uncertainties on the optical constants at millimetre wavelengths. 
As a sanity check, we calculate the flux in the ATCA band using a power-law extrapolation of the SED. In the far-IR to mm regime, the model emission declines as $\lambda\uma{-2.51\pm0.16}$\footnote{Note that, at longer wavelengths, the emission seems to falls-off faster than a simple power-law.}. This results in a flux equal to $1.6^{+0.4}_{-0.3}$\,mJy, which is
within $\sim3\,\sigma$ compatible with the measured flux of $0.72\pm0.25$\,mJy. 
Thus, the model remains valid for the longest wavelength observations (coldest emitting particles),
showing that the ring resolved in the near-IR with the HST is compatible with the spatial distribution of the largest grains, at least at first order.

A limit of our approach is the assumption that the grain size distribution is the
same everywhere in the disk, while it is known that only the smallest
grains can populate the outer disk due to radiation pressure.  
The maximum size and power law slope of the dust size distribution should vary with the distance from the star \citep[e.g. Fig.\,4 in][]{augereau2001}, whereas the
model keeps both large grains in the outer disk and a constant slope
for the size distribution.
%Therefore, one possibility is that ATCA tracks more closely the parent-belt population because it is sensitive to larger grains.
%We indeed observe evidence for such a size differentiation: we convolve the extrapolation of our model with the ATCA beam to obtain a FWHM of [TBC]. This is larger than the observation FWHM (7.8$\arcsec$).
%
Such a radial variation of the size distribution is suggested by
a comparison of a synthetic image at 70\um\ with the 
PACS 70\,\um\ map (Sec.\,\ref{sub:resolve}). The synthetic image
appears broader than the observed one with a FWHM of 7.68$\arcsec$,
versus 6.44$\arcsec$ for the PACS image. 
Our interpretation is that the PACS band reveals larger grains that 
experience less the force due to the radiation pressure and thus are more
bound to the parent belt.
Assuming all grains are equally distributed results in
putting too many large grains in the outer disk: a more complex model should account for the
true dynamics of the particles.

\subsection{Emission from inside the belt?}
The region of the disk interior to 75\,AU cannot be reliably assessed
from the NICMOS data because of the coronagraph and artefacts from the
PSF-substraction interior to 1.5$\arcsec$ (78 AU).
Therefore we cannot exclude the existence of an additional dust
component interior to the main belt, that would be invisible in the
surface density profile we use.
If micron-sized grains populate the inner disk, they would radiate
mainly in the mid-IR, which might affect the results of our study.
From the Gemini/T-ReCS images of \citet{Chen08}, we estimate that less
than $\sim$15\% of the light emitted by the disk at 18.3\,\um\
originates from a region interior to 1$\arcsec$ \citep[Fig.\,2
in][]{Chen08}.

We perform a simple test of the robustness of our results. Suppose that a belt of black body dust particles is orbiting the star
at a distance $r < 1\arcsec$.  We calculate a black body spectrum than
peaks at 18.3\,\um\ ($T~=~150$\,K) and we adjust it such that its
maximum flux represents 15\% of the 53\,mJy attributed to the disk at
that wavelength. Then we subtract the black body flux from the
observed SED of Tab.\,\ref{tab:photometry} to simulate an observation
of the system excluding the hypothetical inner component.  We
reproduce the approach of Sec.\,\ref{sec:ism} with this new SED.
Interestingly the results of the Bayesian analysis are not significantly 
impacted.  To account for the deficit of flux at mid-IR wavelengths,
only the minimum grain size needs to be slightly adjusted to eliminate
the smallest grains: $a\dma{min} = 1.0\,$\um\ and $\kappa = -3.43$,
while we have $a\dma{min} = 0.89\,$\um\ and $\kappa = -3.42$ for our
best fit (Tab.\,\ref{tab:dustmodels}). The composition remains
unchanged for the best model ($\chi^2\dma{r} = 1.52$) and no change is
induced that would improve the consistency of the model with the
scattered light flux (Sec.\,\ref{sec:scattering}).
Consequently we cannot reject the possible existence of an inner
component to the disk, but none of our conclusions would suffer from
its existence.
\section{Grain dynamics and ice origin}\label{sec:timescales}
\subsection{Blowout size}
Grains in the \hd\ debris disk
%are essentially subject to a balance between
%the central star gravity and radiation pressure.  If, for a given
%grain composition, 
for which radiation pressure overcomes gravity are put on unbound
orbits and eventually ejected from the system on a short time
scale \citep[see for instance][and ref. therein]{Krivov2010}.  The
balance between radiation pressure ($F\dma{P}$) and stellar gravity
($F\dma{G}$) forces is evaluated with the distance independent, but
size-dependent, $\beta\dma{pr}$ factor, given by
\begin{eqnarray}
	\bpr = \frac{\vert F\dma{P}\vert}
		{\vert{F\dma{G}}\vert}
		= \frac{\sigma}{m}\frac{
				 \langle Q\dma{pr}\rangle
				 L_{\star}}
		{4 \pi c GM\dma{\star}}
  \label{eq:beta_pr}
\end{eqnarray}
\begin{figure}[!h]
  \begin{center}
    \includegraphics[angle=0,width=0.95\columnwidth,origin=bl]{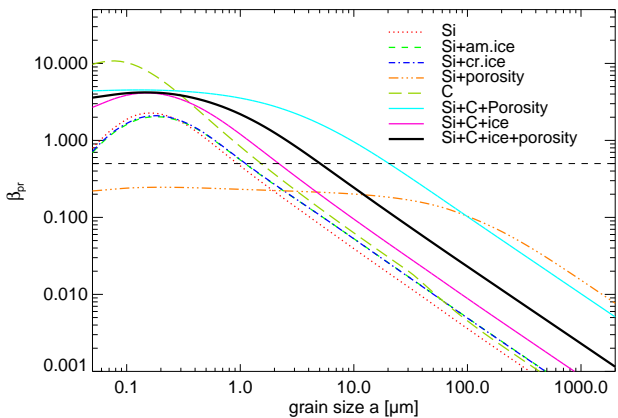}
     \caption{$\bpr = {\vert F\dma{P}\vert} / {\vert F\dma{G}\vert}$ as a function of grain radius. The horizontal dashed line denotes the limit between bound ($\beta\,<\,1/2$) and unbound ($\beta\,>\,1/2$) grains. }
    \label{fig:beta_pr}
  	\end{center}
\end{figure}
\noindent where $\langle Q\dma{pr} \rangle$ is the pressure efficiency
averaged over the stellar spectrum, $m$ the grain mass, $\sigma$ the
grain geometrical cross section, $c$ the speed of light, $G$ the gravitational constant, and $L\dma{\star}$ and $M\dma{\star}$ documented in Tab.\,\ref{tab:stardisk}.
Assuming that the grains are spherical, Eq.\,\ref{eq:beta_pr} yields a
critical size below which the grains originating from the parent belt
are ejected from the system: ${a\dma{blow}=a(\bpr = 1/2)}$ (assuming
circular orbits for the parent bodies).
%

%
%\begin{eqnarray}
%	\bpr = \frac{3L_{*}Q_{pr}}
%			{16 \pi G M_{*} c a \rho}
 % \label{eq:beta_sph}
%\end{eqnarray}
%
%where $a$ is the grain size and $\rho$ the grain density.
In Fig.\,\ref{fig:beta_pr}, we represent the $\beta\dma{pr}$ ratios as
a function of grain radius for different compositions.  Our best
4-component grain model has a blowout size $a\dma{blow} = 4.83$\,\um,
that is 5 times larger than the inferred minimum grain size in the
system.  Models implying no porosity nor carbon, on the other hand, have
blowout sizes of the order of the inferred value of $a\dma{min}$, more
consistent with the theoretical argument that grains below the blowout
size should be expelled from the system.
The response of silicate grains to the stellar irradiation and gravity
fields is only slightly affected by the presence of ice, leading to
similar shape and moderated values for $\beta\dma{pr}(a)$.  Organic
refractory grains on the other hand are highly sensitive to the
radiation pressure and they can reach high $\beta\dma{pr}$ values.
Fig.\,\ref{fig:beta_pr} shows that porosity has the effect of flattening
the $\beta\dma{pr}(a)$ function, pushing the critical value
$\beta\dma{pr}=1/2$ to larger grains.  Hence the addition of porosity
to a mixture of carbon and silicates produces a strong effect:
$\beta\dma{pr}$ not only reaches a high value at its maximum due to
carbon, but the $\beta\dma{pr}(a)$ function decreases slowly such that
the critical value $\beta\dma{pr} = 1/2$ is reached only for large
grain sizes.  

Decreasing the amount of carbon in the grains 
(it was imposed somehow arbitrarily, see Sec.\,\ref{sec:ism})
would mechanically decrease their blowout size. Pure silicate grains
with the same porosity and ice content have $a\dma{blow} = 0.89$\,\um,
but they do not provide good fits to the SED.
We computed additional grid of models with alternative
$v\dma{C}/v\dma{Si}$ values.
% ($v\dma{C} / (v\dma{C}+v\dma{Si}) = 0.0, 0.1, 1.0$). 
The results confirm the conclusion of Sec.\,\ref{sec:si+acar}: all
Si+C+H$_2$O+porosity models lead to similarly good fits to the SED, as long as a
sufficient fraction of carbon is considered.  For instance, if
silicates represent only 10\% of the C+Si mixture, then $v\dma{ice}$
remains at 0.65 and the porosity is reduced to 55\%, with
$M\dma{dust} = 0.043\,M\dma{\oplus}$, $a\dma{min} = 0.89\,$\um\ and
$\chi^2\dma{r} = 1.66$. All these values remain within the uncertainties of
the original model.
Models involving carbon-poor grains do not provide good fits to the
SED, but as long as $v\dma{C}/v\dma{Si} \gtrsim 0.5$, the results
remain unchanged.
% Il faut au moins un peu de carbone : 10% c'est pas assez. 1/3 carbon ça commence à être bien chi2 = 1.8
%
In summary, the discrepancy between $a\dma{min}$ and $a\dma{blow}$
does not lie on the hypothesis we made on the composition of the
grains. 

Our estimation of $a\dma{blow}$ is based on the assumption
that the grains are spherical and homogeneous, but the actual response
of inhomogeneous aggregates to radiation pressure cannot be that
simple.
\citet{Saija2003} for instance show that the fluffiness of the grains
tends to decrease their radiation pressure cross-section, as if
aggregates behaved like their elemental constituents. This would make
grains below the theoretical blowout size less sensitive to radiation
pressure, and allow them to remain bound to the star. Thus the
apparent inconsistency may be interpreted as another signpost of the complex
structure of the grains. We emphasize that fixing $a\dma{min}$ as
equal to $a\dma{blow}$ calculated with the (EMT-)Mie theory would not
be a relevant approach when dealing with porous grains.

Another possibility is that grains
are not in dynamical equilibrium (i.e. an imbalance between production
and ejection), following for instance a major collisional event, such
as an asteroid break-up.  An excess of dust grains from all sizes
would have been produced, including grains below the blowout
size. These grains could have survived destructive collisions and radiation pressure
blowing for long enough to be observed.  Of course, observing such an
event is unlikely, because dust typical lifetime is only of the
order of 1,000 years, which is small with comparison to the age of the
star (see Sec.\,\ref{discu:timescale}).  But this "rare event
hypothesis" would be consistent the scenario proposed by
\citet{Chen08} to justify the brightness asymmetry they claim to
observe in thermal light.

\subsection{Time scales and origin of the ice}
\label{discu:timescale}

%\textit{Time scales} - 
Collisions occurring between grains in a debris disk are a key process
as they generally result in the production of smaller dust
particles, either by fragmentation of the target, or 
excavation. Any phenomenon longer than the characteristic collisional time scale will be considered negligible in the
evolution of the dust population.
%We use the approximation of \citet{backman_paresce} to estimate the collisional time scale (Eq.\,\ref{eq:tcol}).
Based on the principle that any particle will cross the disk midplane
twice during one orbit, \citet{2007A&A...472..169T} proposed a
simplified relation for the collisional time scale, valid for the
smallest grains: $t\dma{col} = t\dma{orbit}/\tau\dma{\perp}\uma{geo}$,
with $t\dma{orbit}$ the orbital period at the surface density peak,
and $\tau\dma{\perp}\uma{geo}$ the geometrical vertical optical
depth. Here the orbital period at 89.5\,AU is $\sim 850$\,years,
leading to $t\dma{col} \sim 7600$\,years for the best model.  This is
more than three orders of magnitude smaller than the age of the system and
two orders of magnitude smaller than the characteristic time-scale for
Poynting-Robertson drag: $t\dma{pr} = 0.1$\,Myr for $\beta\dma{pr} =
1/2$ grains \citep{wyatt2008}.  The evolution of the \hd\ debris disk
is clearly dominated by collisions requiring a mechanism to
generate dust on a $10\uma{3}-10\uma{4}$\,years characteristic time
scale.

Attempts to model time-dependent chemistry with ProDiMo
reveals that the grain water ice mantle would be photo-evaporated in a few hundred years (for $\sim 1\,$\um\
grains), to a few thousand years (for $\sim 10\,$\um\ grains), a similar (or shorter) time scale to the dust mutual collision,
requiring as well an ice reprocessing mechanism.
%{\bf [TBC: discuss time dependent vs equilibrium chemistry]}
This provides a consistent view that the icy dust grains are
maintained through collisions in a reservoir of icy planetesimals.
%textit{Presence of ice} - 
%{\bf [Introduce the general issue of ice in circumstellar disks, Snowball phase: \citet{Xie:2010fk}]}
We highlight that, while detecting the mid-infrared spectral features
in the emission of a debris disk is very challenging -- unambiguous
detection have only been made in brighter circumstellar disks (see,
e.g. Malfait et al. 1999) -- the sole analysis of the continuum SED alone can
provide clues on the amount of ice in dust disks, thus on the nature
of the parent bodies, often referred to as the leftovers of planet
formation.

In the Solar System Kuiper Belt, the study of icy planetesimals, Kuiper Belt Objects (KBOs)
in particular, is a prolific topic \citep[for a recent review
see e.g.][]{jewitt2010}.  The presence of ice, together with
the porous nature of the objects, is inferred from estimates of
their densities.  Unambiguous detection of water ice has been made
possible by the spectroscopic near-IR signature of crystalline ice, for
instance by \citet{Delsanti:2010fk} who completed their study of the
plutino Orcus by radiative transfer considerations concluding that it
contains a mixture dominated by amorphous and crystalline
water ice.  Confronted with space weathering, crystalline ice should be
amorphidized in a short timescale. This implies that ice needs to be
stored (and possibly produced) in the interior of the object before
being brought to the surface, possibly through cryovolcanic events.
The nature of water ice in KBOs remains unclear, but the vast majority
of these objects are too small to experience such geological
processes \citep[][ and references therein]{2011A&A...528A.105D}.
Thus KBOs covered by crystalline ice can only constitute particular
examples, and the water ice covering the collisionally produced dust
in the Solar System is more likely amorphous.  The young age of \hd\
allows the possibility that primordial crystalline ice has survived in
its debris disk.  Our study tilts in favor of the amorphous nature of that
ice, suggesting that either ice was predominantly incorporated in
an amorphous form in \hd's Kuiper Belt objects, or that the
amorphization timescale is shorter than $\sim$10 Myr in this distant
belt.

\section{Discussion}
\label{sec:discuconclu}
The properties of the \hd\ debris disk can be compared to those of the disks surrounding two of its co-moving "sister" stars, \bp\ (A3V, 1.7 $M\dma{\odot}$) and \object{AU~Microscopii} (M1V, 0.5 $M\dma{\odot})$, which 
formed in the same molecular cloud from the same material and are coeval. Of course comparisons are limited by the specific models used to study each object.
The relatively large porosity found here should be put in the context
of the grain properties found for \mic\ \citep{graham07, fitz07}.  
Assuming a Si+C+H$_2$O mixture, only highly porous grains ($\mathcal{P} = 80 \%$), reproduces the scattered light profiles (exhibiting strong forward scattering), the high polarization and the thermal SED.
In that model, the Mie theory remains valid to represent the scattering properties of porous grains, because the grains filling the outer disk are very small and stay in the Rayleigh regime.
A total dust mass of $0.01 M\dma{\oplus}$ is found, that essentially reflects the contribution from the inner mm-sized grains, concentrated in a $\sim$40 AU belt.
A similar porous grain model was used by \citet{LiGreenb} to model \bp\, yielding a dust mass $\sim$0.4 $M\dma{\oplus}$, contained in the $\sim100$ AU-wide outer disk.
Overall, "cold coagulation-like" dust models prove effective at reproducing observations of these young debris disks. 
A trend appears for increasing dust mass with spectral type as could be intuitively expected, but despite this difference of scale, the three stars are orbited by a similar material.

Given the young age of these objects, we are investigating the
conditions holding within disks just after the planetary embryos may have formed,
the gas dispersed, and possible accretion onto outer giant planets happened.
To explain the belt-like structure of the \hd\ disk, the most straightforward explanation is to invoke the dynamical influence of a planet inside the belt.
The \citet{1980AJ.....85.1122W} analytical criteria gives the width of the chaotic zone a planet mean-motion resonances would create.
Assuming circular orbits for the planetesimals, as suggested by the sharpness of the belt, this yields a range of possible values for the planet semi-major axis depending on its mass.
Assuming the inner edge of the disk is located at 78 AU from 
the star, the yet undiscovered planet would lie from 73 AU 
for a Neptune-mass planet, down to 59 AU for a 10 Jupiter-mass 
planet.

The material at that stage appears as very hydrated providing a
considerable reservoir of water in the outer system. 
The existence of such a reservoir 
is hypothesized in scenarios of the Solar System formation to
explain the unexpectedly large water content on Earth for
instance \citep{2000M&PS...35.1309M}.
Indeed, terrestrial planets at this age (12 Myr) are
likely too hot to hold on to water and later impacts of icy comet-like bodies from the outer regions
are a possible mechanism for delivering water and volatiles to planets in the habitable zone.
Thus, not only \hd 's debris disk reveals that the system has already grown planetary embryos -- including maybe the bricks of 
rocky planets in the inner disk -- but it likely holds the conditions that were prevailing before the Solar System Late Heavy Bombardment that depleted the Kuiper Belt and determined terrestrial planets history.
%________________________________________________________________
\section{Summary and conclusions}
\label{sec:conclu}
We have presented new \her/PACS and ATCA imaging of the \hd\ debris
disk, a 12\,Myr old star belonging to the \bp\ moving group. The 70
and 100\,\um\ images as well as the 3.2 millimetre observation reveal a marginally resolved extended
emission. Fluxes in the three PACS photometric channels were derived
and complemented by continuum fluxes obtained from PACS spectroscopic
observations. 
Together, these data
provide precise information on the continuum thermal emission of the debris
disk, but detailed models such as the one developed in this study 
are required to interpret the precise nature
of the emitting material. Degeneracies between the grain properties
and disk's spatial structure are broken in our model by making use of a
reprocessed HST/NICMOS coronagraphic image revealing that the dust is
bound to a narrow belt peaking at 89.5\,AU from the central star. 
We focus our study on
the dust properties (composition, size distribution) using a radiative
transfer code, and run a large grid of debris disk models for
subsequent statistical (Bayesian) analysis. We conclude that the
thermal light spectrum of the disk is well reproduced assuming a
mixture of silicates, carbonaceous material and amorphous ice, as well
as vacuum to mimic a significant porosity.  The observed SED can be
very reliably reproduced by a population of grains larger than $\sim
1\,$\um, which follows a power-law distribution matching the
ideal collisional equilibrium case and represents $\sim$0.05\,$M\dma{\oplus}$ ($1.5\times 10^{-7} M_{\odot}$) in grains with sizes up to 1
mm.
% No upper cut-off value shall be
%accounted in that size distribution, as it is supposed to be valid for
%arbitrarily large grains despite the lack of observable signature upon
%a few millimetres. 
Extending the size distribution to the parent bodies of the observed
dust grains leads
%The mass scales as $M \propto a\dma{max}^{1/2}$ leading 
to a mass of $\sim 50 M\dma{\oplus}$ for planetesimals up to 1\,km.
The composition we infer for the dust grains must be a good
representation of a considerable reservoir of material hidden in
unseen planetesimals.

In this study, we attempted to use Herschel/PACS spectroscopy to detect for the first time far-infrared emission lines in the debris disk.
The non-detections we obtain were used to investigate the gas content of the debris disk 
with our photochemical disk model ProDiMo based on the dust model found with GRaTer.
The main uncertainty lies in the amount of PAHs in the disk.
The current non-detections of [OI] and [CII] lines alone do not
provide unambiguous upper limits to the gas content.  
Only in the case of a high PAH abundance, we can set an upper limit
to the gas
mass of $\sim 17 M_{\oplus}$ (one Neptune mass).  Coupling with other
tracers, in particular with the CO lines accessible to ALMA, offers
much better prospects to reach lower limits on the low gas content of
debris disks. However, steady-state chemistry is likely not ideally
suited to debris disk modeling, and in future studies more appropriate
prescriptions, time-dependent chemistry in particular, will be explored. 

We note that both the low albedo estimated from the scattered light
images and the dust dynamics are not well modelled by spherical grains filled with vacuum, and that the grains are likely
complex fluffy aggregates. Though their thermal behavior is correctly
described by the EMT-Mie theory, the scattering of light at short
wavelengths and their sensitivity to radiation pressure require a more
accurate description of the exact shape of the aggregates.  Our
statistical analysis highlights that ignoring the porosity of the
grains or setting the minimum grain size as equal to the blowout size
would result in incorrect interpretation of the observations, in
particular because the grain opacity is strongly dependent on the exact grain
composition.

We stress some necessary limitations of the models. First, we use a
grain size distribution that is independent of the distance to the
star, whereas it is known that only the smallest grains can populate
the outer parts of the disk under the effect of radiation pressure. In
this paper we discussed clues that the observed scattered light
profile may not be a perfect representation of that seen in thermal
light, by Herschel or ATCA. The long wavelength emission emanating
from the largest grains might originate from an even narrower belt.
Further studies will need to account for the radial differentiation of grain sizes 
with the objective to fit altogether resolved images from the near-IR to the millimetre domain.
Secondly, the Dohnanyi-like power-law size distribution is known to be
an insufficient description; if the real distribution implies wavy
patterns as proposed by \citet{2007A&A...472..169T}, then adjusting a
power-law likely puts too many/too few small grains in the
disk. Dynamical models are required to understand better the general
outcome of the collisional evolution of the disk.
%__________________________________________________________________
\begin{acknowledgements}
We wish to thank Paul Smith and Karl Stapelfeldt for their helpful
comments regarding the Spitzer/MIPS-SED data, and Johan Olofsson for
reducing the Spitzer/IRS spectrum. We are grateful to the anonymous referee who suggested improvement to the manuscript. We also thank the Programme
National de Plan\'etologie (PNP) and the CNES for supporting part of
this research.  JL and JCA thank the French National Research Agency
(ANR) for financial support through contract ANR-2010 BLAN-0505-01
(EXOZODI).
This study is based, in part, on observations made with the NASA/ESA Hubble Space Telescope, obtained at the Space Telescope Science Institute (STScI), which is operated by the Association of Universities for Research in Astronomy, Inc., under NASA contract NAS 5-26555. These observations are associated with program \#\,10177. Support for program \#\,10177 was provided by NASA through a grant from the STScI.
\end{acknowledgements}

%________________________________________________________________
\appendix
\section{Statistical analysis of the results}
\label{sec:statistics}
The strength of the GRaTer code, in comparison with an optimization algorithm that would provide a unique best-fit solution and possibly miss local minima, is that all the models are stored enabling fast and easy post-processing of the results.

To find the best solution in the grid of models, we use a classical least-square minimization approach 
with $n\dma{free} = $ 3 to 4 free parameters depending on the grain composition and $N = 48$, 
which results in 43 to 45 degrees of freedom (dof) (Tab.\,\ref{tab:dustmodels}).
The free parameters are $\kappa$, $a\dma{min}$ and $M\dma{dust}$, 
and depending on the situation $v\dma{C}$, $v\dma{ice}$ or $\mathcal{P}$. 
$M\dma{dust}$ is handled separately because it is automatically adjusted to reproduce the surface density profile in the code. 

Let $\vec{a}$ be the vector of parameters. Here
$\vec{a}~=~(\kappa,a\dma{min},v\dma{ice},\mathcal{P})$.
For each set of parameters, we can define the quantity:
\begin{eqnarray}
	\chi^2(\vec{a}) = \sum_{i=1}^{N} {\frac{(D_{i}-M_{i})^2}{\sigma_{i}^2}}
  	\label{eq:chi2}
\end{eqnarray}
where $D_{i}$ and $M_{i}$ are the observed and modeled values, respectively, and $\sigma_{i}$ is the measured uncertainty.
For each $\chi^2$ value, we can define the likelihood of the data, \ie\ the probability of the data given the parameters of a model. 
This assumes the errors are distributed normally and are uncorrelated.
\begin{eqnarray}
	P(\vec{D} | \vec{a}) = \frac{1}
%		{\prod\limits_{i=1}^{N}{(2\pi)^{1/2}\sigma_{i}}}
	{C}\,
		{\exp{\left(\frac
					{-\chi^2(\vec{a})}
					{2}
					\right)}
		}
\end{eqnarray}
where C is a normalization constant and $D = \lbrace D(i)\rbrace\dma{i=1\ldots N}$.
In order to evaluate the effect of each individual parameter $a_i$ on the quality of the fit, we implement a Bayesian inference method. 
We define the posterior probability of $\vec{a}$ as the probability distribution of the parameters $\vec{a}$ given the data $\vec{D}$. 
It has to be noted that this formulation implicitly assumes that the model is suited to describe the observation.
Bayes' theorem says that this quantity can be written:

\begin{eqnarray}
	P(\vec{a}| \vec{D}) = \frac{P(\vec{a})P(\vec{D} | \vec{a})}
		{P(\vec{D})}
\end{eqnarray}
where $P(\vec{D})$ is the prior probability of the data under all possible hypothesis (a normalization constant in practice), and $P(\vec{a})$ is the prior probability distribution of the parameters $\vec{a}$. 
Here we use uniform priors for all the parameters, meaning that we do not favor any realization of the parameters prior to obtaining the data.
That fundamental theorem provides a simple way to independently assess the parameters of a model. 
Indeed, the marginal probability of each parameter $a\dma{i}$ can be computed 
regardless of the values of the other parameters $\vec{a\dma{j \ne i}}$:

\begin{equation}
	P(a\dma{i} | \vec{D}) = 
			\int\limits_{a\dma{0}}\ldots
			\int\limits_{a_{i-1}}
			\int\limits_{a_{i+1}}\ldots
			\int\limits_{a\dma{i\dma{max}}}
			{P(\vec{a}\dma{j \ne i}	
			| \vec{D} ) da\dma{0} \ldots da\dma{i\dma{max}}}
\end{equation}

Then we obtain a probability law as a function of the realization of each individual parameter.
Finally, we estimate the parameters by calculating the mathematical expectation 
and variance. These are the values presented in Tab. \ref{tab:dustmodels} where the uncertainties are 3$\sigma$.
\begin{eqnarray}
E({a\dma{i}}) &=& \int\limits_{a_{i_{min}}}
										^{a_{i_{max}}}
										{a\dma{i} P(a\dma{i} | \vec{D}) }\\
\sigma^2(a\dma{i}) &=& E({a\dma{i}\uma{2}}) - E({a\dma{i}})\uma{2} 
\end{eqnarray}

%For example, the probability distribution of the ice fraction can be obtained by marginalizing over $\mathcal{P}$, $\kappa$ and $a_{min}$: 
%
%\begin{eqnarray}
%	P(V_{ice} | \vec{D}) = 
%		{\int{
%		\int{
%		\int{P(\mathcal{P}, a_{min}, \kappa | \vec{D}) {d\mathcal{P}} {da_{min}} {d\kappa}}
%			}
%		}
%	}
%\end{eqnarray}	
%
%Then we can plot a probability curve projected on $V_{ice}$ by summing over all the other parameters for each $V_{ice}$. 
%To study the likelihood of different compositions, we use 2D-probability maps, where we integrate the probability over the couples ($a_{min}$, $\kappa$) .
%
%In the following, we consider several mixtures of materials and we present normalized probability curves for the individual parameters. 
%\input{atca}

\section{Gas line fluxes}\label{sec:tab_gas}
\begin{table*}[h!tpb]
\begin{center}
  \caption{Gas line fluxes. The gas to solid mass ratios assume a
    solid mass of 0.164 M$_{\oplus}$. The PACS 3-$\sigma$ upper limits
    do not include the 30\% flux calibration uncertainty.}
\label{tab:gas}
\begin{tabular}{cccccccrrrrr}
  \hline
  \noalign{\smallskip} 
  Gas-to-Solid & $\mathrm{[O~{\sc I}]}$~63$\mu$m & $\mathrm{[O~{\sc I}]}$~145$\mu$m  & $\mathrm{[C~{\sc II}]}$~158$\mu$m & CO $J$=3 $\rightarrow$ 2 & CO $J$=2 $\rightarrow$ 1\\
  mass ratio & (W m$^{-2}$) & (W m$^{-2}$) & (W m$^{-2}$) & (W m$^{-2}$)& (W m$^{-2}$)\\
  \noalign{\smallskip}   
  \hline
  \noalign{\smallskip}   
  & \multicolumn{3}{c}{Herschel-PACS (3$\sigma$)} & &\\
  \noalign{\smallskip}   
  \cline{2-4}
  \noalign{\smallskip}   
  & 9.8$\times$10$^{-18}$& 8.5$\times$10$^{-18}$ & 7.0$\times$10$^{-18}$ & - & -\\
  \hline
  \noalign{\smallskip} 
  &\multicolumn{5}{c}{ProDiMo predictions with $f_{\mathrm{PAH}}$=10$^{-5}$}\\
  \noalign{\smallskip} 
  1000  &2.9$\times$10$^{-18}$&1.6$\times$10$^{-18}$&2.6$\times$10$^{-20}$&1.7$\times$10$^{-19}$&5.6$\times$10$^{-20}$\\
  100 &6.7$\times$10$^{-19}$&1.2$\times$10$^{-20}$&4.1$\times$10$^{-20}$&1.5$\times$10$^{-19}$&5.3$\times$10$^{-20}$\\
  10       &2.1$\times$10$^{-18}$&8.2$\times$10$^{-21}$&2.2$\times$10$^{-19}$&1.3$\times$10$^{-19}$&3.8$\times$10$^{-20}$\\
  1        &2.4$\times$10$^{-18}$&4.0$\times$10$^{-20}$&6.2$\times$10$^{-19}$&3.4$\times$10$^{-22}$&6.1$\times$10$^{-23}$\\
  0.1      &6.4$\times$10$^{-19}$&5.0$\times$10$^{-20}$&1.9$\times$10$^{-19}$&1.0$\times$10$^{-23}$&1.6$\times$10$^{-24}$\\
  0.01 &2.8$\times$10$^{-20}$&2.9$\times$10$^{-21}$&1.5$\times$10$^{-20}$&1.1$\times$10$^{-24}$&2.7$\times$10$^{-25}$\\
  \noalign{\smallskip}   
  \hline
  \noalign{\smallskip} 
  &\multicolumn{5}{c}{ProDiMo predictions with $f_{\mathrm{PAH}}$=0.1}\\
  \noalign{\smallskip} 
  1000  &2.7$\times$10$^{-17}$&6.4$\times$10$^{-18}$&4.1$\times$10$^{-20}$&4.8$\times$10$^{-19}$&1.6$\times$10$^{-19}$\\
  100  &7.1$\times$10$^{-18}$&2.0$\times$10$^{-19}$&6.0$\times$10$^{-20}$&3.1$\times$10$^{-19}$&1.0$\times$10$^{-19}$\\
  10       &3.2$\times$10$^{-18}$&1.5$\times$10$^{-20}$&1.4$\times$10$^{-19}$&1.4$\times$10$^{-19}$&3.9$\times$10$^{-20}$\\
  1        &2.4$\times$10$^{-18}$&4.0$\times$10$^{-20}$&4.2$\times$10$^{-19}$&2.6$\times$10$^{-22}$&4.6$\times$10$^{-23}$\\
  0.1      &6.3$\times$10$^{-19}$&4.8$\times$10$^{-20}$&1.8$\times$10$^{-19}$&1.0$\times$10$^{-23}$&1.5$\times$10$^{-24}$\\
  0.01 &2.8$\times$10$^{-20}$&2.9$\times$10$^{-21}$&1.5$\times$10$^{-20}$&1.1$\times$10$^{-24}$&2.7$\times$10$^{-25}$\\
  \noalign{\smallskip}   
  \hline
\end{tabular}
\end{center}
\end{table*}

%__________________________________________________________________
\bibliography{bib}

%__________________________________________________________________

\end{document}